\documentclass[
superscriptaddress,aps,preprintnumbers,amsmath,amssymb,prd,nofootinbib,preprint]{revtex4-1}
\usepackage{graphicx}
\usepackage{epstopdf}
\usepackage{dcolumn}% Align table columns on decimal point
\usepackage{bm}% bold math
\usepackage{hyperref}
\usepackage{color}
\usepackage{amsmath}
\usepackage{cancel}

\begin{document}

%%%%%%%%%%%%%%%%%%%%%%%%%%%%%%%%%%%%%%%%%%%

\def\a{\alpha}
\def\b{\beta}
\def\c{\varepsilon}
\def\d{\delta}
\def\e{\epsilon}
\def\f{\phi}
\def\g{\gamma}
\def\h{\theta}
\def\k{\kappa}
\def\l{\lambda}
\def\m{\mu}
\def\n{\nu}
\def\p{\psi}
\def\q{\partial}
\def\r{\rho}
\def\s{\sigma}
\def\t{\tau}
\def\u{\upsilon}
\def\v{\varphi}
\def\w{\omega}
\def\x{\xi}
\def\y{\eta}
\def\z{\zeta}
\def\D{\Delta}
\def\G{\Gamma}
\def\H{\Theta}
\def\L{\Lambda}
\def\F{\Phi}
\def\P{\Psi}
\def\S{\Sigma}

\def\o{\over}
\def\beq{\begin{align}}
\def\eeq{\end{align}}
\newcommand{\gsim}{ \mathop{}_{\textstyle \sim}^{\textstyle >} }
\newcommand{\lsim}{ \mathop{}_{\textstyle \sim}^{\textstyle <} }
\newcommand{\vev}[1]{ \left\langle {#1} \right\rangle }
\newcommand{\bra}[1]{ \langle {#1} | }
\newcommand{\ket}[1]{ | {#1} \rangle }
\newcommand{\EV}{ {\rm eV} }
\newcommand{\KEV}{ {\rm keV} }
\newcommand{\MEV}{ {\rm MeV} }
\newcommand{\GEV}{ {\rm GeV} }
\newcommand{\TEV}{ {\rm TeV} }
\newcommand{\1}{\mbox{1}\hspace{-0.25em}\mbox{l}}
\newcommand{\headline}[1]{\noindent{\bf #1}}
\def\diag{\mathop{\rm diag}\nolimits}
\def\Spin{\mathop{\rm Spin}}
\def\SO{\mathop{\rm SO}}
\def\O{\mathop{\rm O}}
\def\SU{\mathop{\rm SU}}
\def\U{\mathop{\rm U}}
\def\Sp{\mathop{\rm Sp}}
\def\SL{\mathop{\rm SL}}
\def\tr{\mathop{\rm tr}}
\def\mpl{M_{\rm Pl}}

\def\IJMP{Int.~J.~Mod.~Phys. }
\def\MPL{Mod.~Phys.~Lett. }
\def\NP{Nucl.~Phys. }
\def\PL{Phys.~Lett. }
\def\PR{Phys.~Rev. }
\def\PRL{Phys.~Rev.~Lett. }
\def\PTP{Prog.~Theor.~Phys. }
\def\ZP{Z.~Phys. }

\def\dd{\mathrm{d}}
\def\ff{\mathrm{f}}
\def\BH{{\rm BH}}
\def\inf{{\rm inf}}
\def\ev{{\rm evap}}
\def\eq{{\rm eq}}
\def\SM{{\rm sm}}
\def\Mpl{M_{\rm Pl}}
\def\GeV{{\rm GeV}}
\newcommand{\Red}[1]{\textcolor{red}{#1}}
\newcommand{\TL}[1]{\textcolor{blue}{\bf TL: #1}}

\title{
Implications of Higgs Discovery \\
for the Strong CP Problem %\\
and Unification
}

\author{Lawrence J. Hall}
\affiliation{Department of Physics, University of California, Berkeley, California 94720, USA}
\affiliation{Theoretical Physics Group, Lawrence Berkeley National Laboratory, Berkeley, California 94720, USA}
\author{Keisuke Harigaya}
\affiliation{Department of Physics, University of California, Berkeley, California 94720, USA}
\affiliation{Theoretical Physics Group, Lawrence Berkeley National Laboratory, Berkeley, California 94720, USA}

\begin{abstract}
A $Z_2$ symmetry that extends the weak interaction, $SU(2)_L \rightarrow SU(2)_L \times SU(2)'$, and the Higgs sector, $H(2) \rightarrow H(2,1) + H'(1,2)$, yields a Standard Model quartic coupling that vanishes at scale $v' = \vev{H'} \gg \vev{H}$.  Near $v'$, theories either have a ``prime" sector, or possess ``Left-Right" (LR) symmetry with $SU(2)' = SU(2)_R$.   If the $Z_2$ symmetry incorporates spacetime parity, these theories can solve the strong CP problem. The LR theories have all quark and lepton masses arising from operators of  dimension 5 or more, requiring Froggatt-Nielsen structures.  Two-loop contributions to $\bar{\theta}$ are estimated and typically lead to a
neutron electric dipole moment of order $10^{-27}$e cm that can be observed in future experiments.  Minimal models, with gauge group $SU(3) \times SU(2)_L \times SU(2)_L \times U(1)_{B-L}$, have precise gauge coupling unification for $v' = 10^{10\pm1}$ GeV, successfully correlating gauge unification with the observed Higgs mass of $125$ GeV. With $SU(3) \times U(1)_{B-L}$ embedded in $SU(4)$, the central value of the unification scale is reduced from $10^{16-17}$ GeV to below $10^{16}$ GeV, improving the likelihood of proton decay discovery.  Unified theories based on $SO(10) \times CP$ are constructed that have $H+H'$ in a {\bf 16} or {\bf 144} and generate higher-dimensional flavor operators, while maintaining perturbative gauge couplings. 

\end{abstract}

\date{\today}

\maketitle

\tableofcontents

%%%%%%%%%%%%%%%%%%%
\section{Introduction}
%%%%%%%%%%%%%%%%%%%

In moving towards a UV completion of the Standard Model (SM), the vast majority of work in recent decades has assumed new physics at around the TeV scale. However, the Large Hadron Collider  has discovered a highly perturbative Higgs boson with mass 125 GeV, but no clear evidence for physics beyond the SM.  These results suggest that an unconventional framework of particle physics should be taken seriously, with the SM the correct effective theory to very high energy scales.  Remarkably, the observed value of the Higgs mass results in the SM Higgs quartic coupling vanishing.  
\begin{align}
\label{eq:muc}
\lambda_{SM}(\mu_c) = 0  
\end{align}
at a scale $\mu_c \simeq (10^9 - 3 \times 10^{12})$ GeV~\cite{Buttazzo:2013uya} (see~\cite{Lindner:1988ww,Sher:1993mf,Altarelli:1994rb,Casas:1994qy,Espinosa:1995se,Casas:1996aq,Hambye:1996wb,Isidori:2001bm,Degrassi:2012ry} for earlier works), or even higher if the top mass is below its measured value by more than $2 \sigma$.  In this framework, we take the scale for new physics to be $\mu_c$.   The weak scale is highly fine-tuned; this might result from environmental requirements \cite{Agrawal:1997gf,Hall:2014dfa}, and should not prevent an exploration of this new picture. 

Several ideas for the new physics that lead to the vanishing of the quartic have been proposed.  It could be that a continuous global symmetry of an extended Higgs potential is spontaneously broken at $\mu_c$, leading to the Higgs boson becoming a Pseudo-Nambu-Goldstone Boson (PNGB), with potential arising from loop corrections~\cite{Redi:2012ad}.  Other possibilities include supersymmetry at $\mu_c$ with $\tan \beta = 1$~\cite{Hall:2013eko,Ibe:2013rpa,Hall:2014vga}, and anthropic arguments associated with vacuum instability~\cite{Feldstein:2006ce}.   

In this paper we introduce a new mechanism for physics at $\mu_c$ that leads to (\ref{eq:muc}).  We assume that the entire theory at $\mu_c$ is invariant under an exact $Z_2$ symmetry that interchanges the SM Higgs doublet $H$ with its $Z_2$ partner $H'$.  This $Z_2$ is spontaneously broken by the condensation $\vev{H'} = v'$ at scale $\mu_c$.   $H'$ must be neutral under $SU(2)_L$ because $v' \gg v$, the electroweak scale. $Z_2$ symmetry requires that $H'$ is a doublet under a partner $SU(2)'$ gauge symmetry.  We study the most general $Z_2$-invariant potential for $(H',H)$. In the required limit of $v \ll v'$, the potential possesses an accidental $SU(4)$ symmetry so that the SM Higgs boson is a PNGB at scale $v' \sim \mu_c$, leading to (\ref{eq:muc}) before including the usual SM radiative corrections. Indeed, we find that this mechanism is highly constrained and requires that $SU(2)_L \times SU(2)'$ symmetry breaking is essentially unique and pristine in its simplicity: $H(2,1) + H'(1,2)$.  This simple gauge structure for the Higgs has several key implications.

We assume the SM $SU(2)_L$-doublet fermions are singlets under $SU(2')$, so that the theory contains $q(2,1) + \ell(2,1) + q'(1,2) + \ell'(1,2)$. Two classes of theories then emerge. In the first class $(q',\ell')$ are identified as the SM $SU(2)_L$-singlet fermions, so that $SU(2)'$ contains the right-handed $W$ and we call it $SU(2)_R$.   In the second class $(q',\ell')$ do not have the correct color and hypercharge to be identified as SM states and they form part of a sector that acquires mass from $v'$, which we call the prime sector.

The $Z_2$ symmetry at $v'$ solves the strong CP problem~\cite{'tHooft:1976up} if it is extended to include spacetime parity and if it does not replicate the QCD gauge group.  With parity included, at scale $v'$ we call the discrete symmetry $P_{LR}$ in models with $SU(2)_L \times SU(2)_R$, and $P'$ in models with a prime sector.  The strong CP problem is solved in both classes of theories:  $P_{LR}$ forces the quark mass matrices to be Hermitian, while $P'$ forces the strong CP phase from colored triplets in the prime sector to cancel that from quarks in the SM sector. Quantum corrections may generate a small but non-zero strong CP phase, which we study.

Parity solutions to the strong CP problem have a long history, starting in 1978 when the possibility of Hermitan quark mass matrices in left-right (LR) symmetric models was stressed~\cite{Beg:1978mt,Mohapatra:1978fy}.  However, a viable solution in conventional LR models, where the SM Higgs doublet is incorporated into $\phi(2,2)$ under $SU(2)_L \times SU(2)_R$, is problematic~\cite{Barr:1991qx}. Phases in the Higgs potential lead to a phase in the vacuum, reintroducing the strong CP problem, unless supersymmetry is added~\cite{Kuchimanchi:1995rp,Mohapatra:1995xd,Mohapatra:1996vg}.  Nevertheless, in LR theories, with the non-standard embedding of the SM Higgs into $H_L(2,1) + H_R(1,2)$, a simple solution to the strong CP problem  was discovered by Babu and Mohapatra in 1989~\cite{Babu:1989rb}.  However, it was based on an approximate parity that was softly broken to obtain a large symmetry breaking scale for $SU(2)_R$.  Our starting point is different.  We find that (\ref{eq:muc}) results {\it only} if $Z_2$ is exact and  {\it only} if the SM Higgs is embedded as $H(2,1) + H'(1,2)$, and we follow the implications of this understanding of the Higgs boson mass.

The gauge charges of SM fermions can be simply understood from grand unification, but gauge coupling unification in the SM lacks precision and leads to unacceptable proton decay, unless very large threshold effects are invoked.   Our new understanding of the Higgs mass implies that new physics lies near $\mu_c$ where $\lambda_{SM}$ vanishes.  Could this new physics lead to precise unification?  Running of gauge couplings is substantially altered because there is an additional $SU(2)$ gauge group but, to solve strong CP, the color gauge group is not replicated.  In the class of theories involving a prime sector at $v'$, the new heavy states couple to the color gauge group but not to $SU(2)_L$, so for $v' \sim 10^{9 -13}$ GeV, the SM couplings $g_2$ and $g_3$ do not meet by the Planck scale.  However, in the minimal LR symmetric model with gauge group $SU(3)_c \times SU(2)_L \times SU(2)_R \times U(1)_{B-L}$ (3221), gauge coupling unification occurs with very high precision for $v' \sim 10^{10}$ GeV, right in the region that yields the observed Higgs mass.  Furthermore, the unification scale is of order $10^{16-17}$ GeV.  If the gauge group above $v'$ is $SU(4) \times SU(2)_L \times SU(2)_R$ (422), $v'$ is increased and the unification scale decreases so that planned proton decay searches become more powerful probes of the theory. 

We find that the LR symmetric models may be unified very neatly into $SO(10)$, with matter in {\bf 16}s~\cite{Georgi:1974my,Fritzsch:1974nn}.  
For $\lambda_{SM}(\mu_c)=0$, the Higgs must transform as $H(2,1) + H'(1,2)$ under $SU(2)_L \times SU(2)_R$ and therefore as a {\bf 16} rather than a {\bf 10} of $SO(10)$, greatly affecting the structure of the theory.   $SO(10)$ contains a generator $C_{LR}$ that is charge conjugation together with $SU(2)_L \leftrightarrow SU(2)_R$, so $P_{LR} = CP *C_{LR}$.  Hence the unified theory must have a symmetry $CP \times SO(10)$ broken by the condensation of a field that is both $C_{LR}$ and CP odd, such as a CP odd {\bf 45} 
\begin{align}
\label{eq:45SBintro}
SO(10) \times CP \;\;  \stackrel{ \phi_{45}}{\longrightarrow} \;\;  SU(3) \times SU(2)_L \times SU(2)_R \times U(1)_{B-L} \times P_{LR}.
\end{align}
In the SO(10) theory, the discrete symmetry that ultimately leads to strong CP conservation is CP; P is not defined.

Gauge coupling unification and the $SO(10)$ embedding only work for the LR theories, not for the theories with prime sectors.  In any case, the LR theories have minimal field content and appear more elegant.  However, with the Higgs in $H(2,1) + H'(1,2)$ they do not have Yukawa couplings.  Thus the flavor problem cannot be postponed, and we construct Froggatt-Nielsen type theories~\cite{Froggatt:1978nt}. Flavor is particularly pressing for the heavy quarks and leptons where the corresponding heavy vector fermions cannot be far above $v'$.  

In Section \ref{sec:Z2} we demonstrate that a $Z_2$ with $H(2,1) \leftrightarrow H'(1,2)$ leads to (\ref{eq:muc}). 
In Section \ref{sec:Models} we explore minimal models at scale $v'$ of both LR and prime sector classes, showing explicitly the possibilities for constructing flavor operators.
In Section \ref{sec:CP} we show that, by including parity in the definition of the $Z_2$ symmetry at the scale $v'$, the strong CP problem can be solved in all these models, provided the color group is $Z_2$-invariant.
In Section \ref{sec:gcu} we explore gauge coupling unification in both 3221 and 422 schemes, and draw conclusions for proton decay.
In Section \ref{sec:so10} we construct a variety of $SO(10)$ models, including operators that lead to quark $(u,d)$, charged lepton $(e)$ and neutrino $(\nu)$ masses. We show how $d/e$ mass splittings arise, and how $\nu$ masses become decoupled from $u$ masses.  We find that precision gauge coupling unification is possible even in the presence of the required Froggatt-Nielsen states.

%%%%%%%%%%%%%%%%%%%%%%%%%%
\section{Vanishing Higgs Quartic from a $Z_2$ Symmetry}
\label{sec:Z2}
%%%%%%%%%%%%%%%%%%%%%%%%%%%
In this section we show that the near vanishing of the SM Higgs quartic coupling at a high energy scale $v'$ can be explained by a $Z_2$ symmetry spontaneously broken at $v'$.

%%%%%%%%%%
\subsection{Vanishing quartic coupling}
%%%%%%%%%%
We introduce a $Z_2$ symmetry under which the Higgs field $H$ and its partner $H'$ are exchanged. The renormalizable potential of $H$ and $H'$ is given by
\begin{align}
\label{eq:potential}
V(H,H') = - m^2 (H^\dagger H + H'^\dagger H') + \frac{\lambda}{2} (H^\dagger H + H'^\dagger H')^2 + \lambda' H^\dagger H H'^\dagger H' .
\end{align}
We assume that the mass scale $m$ is much larger than the electroweak scale. With $m^2$ positive, the $Z_2$ symmetry is spontaneously broken and $H'$ acquires a large vacuum expectation value of $\vev{H'} = v'$, with $v'^2 = m^2/\lambda$. After integrating out $H'$ at tree-level, the Low Energy potential in the effective theory for $H$ is
\begin{align}
\label{eq:potentialLE}
V_{LE}(H) = \lambda' \; v'^2  \; H^\dagger H - \lambda' \left(1  + \frac{\lambda'}{2 \lambda} \right) (H^\dagger H)^2 .
\end{align}
In order to obtain the hierarchy $\vev{H} = v \ll v'$, it is necessary that $\lambda' \ll 1$. After this fine-tuning, the quartic coupling of the Higgs $H$, $\lambda_{\rm SM}$, also vanishes.

The vanishing quartic can be understood by an accidental $SU(4)$ symmetry under which $(H, H')$ is a fundamental representation. For $\lambda' = 0$, the potential in Eq.~(\ref{eq:potential}) is manifestly $SU(4)$ symmetric.
 After $H'$ obtains a vacuum expectation value, the Standard Model Higgs is understood as a Nambu-Goldstone boson with a vanishing potential.

Below the scale $v'$, quantum corrections from SM particles renormalize the quartic coupling, and it becomes positive.
From the perspective of running from low to high energies, the scale at which the SM Higgs quartic coupling vanishes, $\mu_c$, is to be identified with $v'$ as in (\ref{eq:muc})  $v' \simeq \mu_c$.  The value of $\mu_c$ depends on uncertainties in the top quark Yukawa coupling, and is listed in Table~\ref{tab:vp},
following the calculation of Ref.~\cite{Buttazzo:2013uya}.

\begin{table}[tb]
\caption{The value of $\mu_c$ for several top quark masses. Here we use $m_h = 125.1$ GeV and $\alpha_3 (m_Z) = 0.1184$.}
\begin{center}
\begin{tabular}{|c|c|c|c|c|c|} 
\hline 
$m_t$ (GeV) & 171.9 & 172.5 & 173.1 & 173.7 & 174.3 \\
$\mu_c$ (GeV) & $3 \times 10^{12}$ & $1\times 10^{11}$ & $2\times 10^{10}$ & $3 \times 10^9$ & $8 \times 10^8$ \\ \hline
\end{tabular}
\end{center}
\label{tab:vp}
\end{table}%

Although the scale $v'$ is much smaller than the Planck scale and the typical unification scale, the theory is no more fine-tuned than the Standard Model because of the $Z_2$ symmetry.
The required fine-tuning is
\begin{align}
\frac{m^2}{\Lambda^2} \times \frac{v^2}{m^2} = \frac{v^2}{\Lambda^2},
\end{align}
where the first factor in the left hand side is the fine-tuning to obtain the scale $m$ much smaller than the cut off scale $\Lambda$, and the second one is the fine-tuning in $\lambda'$ to obtain the electroweak scale from $m$. The total tuning is the same as in the Standard Model, $v^2 / \Lambda^2$, and may be explained by environment requirements \cite{Agrawal:1997gf,Hall:2014dfa}.

%%%%%%%%%%%%%%%
\subsection{The symmetry breaking sector at scale $v'$}
%%%%%%%%%%%%%%
Since $H'$ obtains a large vacuum expectation value, it cannot have the same Standard Model gauge charges as $H$. We must introduce an additional $SU(2)'$ gauge symmetry, under which $H'$ is charged. The $Z_2$ symmetry exchanges the two $SU(2)$ gauge symmetry groups: $SU(2)_L \leftrightarrow SU(2)'$, for example
\begin{align}
\label{eq:H+H'}
H(2,1) \;\; \leftrightarrow \;\; H'(1,2).
\end{align}
For our $Z_2$ understanding of $\lambda_{SM}(\mu_c) = 0$, in Appendix A we show that $SU(2)'$ symmetry breaking at scale $v'$ can occur {\it only} by the vev of $H'(1,2)$. Additional scalars multiplets may exist at scale $v'$ if they have no vev, but they lead to extra fine-tuning and we do not add them.  Furthermore, in Appendix~\ref{sec:Higgs sector} we show that the low energy field that breaks $SU(2)_L$ {\it must} lie dominantly in $H(2,1)$, and hence we similarly do not add extra light scalar multiplets non-trivial under $SU(2)_L$.   
In this paper, $SU(2)_L \times SU(2)'$ symmetry breaking is accomplished by $H(2,1)+H'(1,2)$ alone, although the quantum numbers of $H$ and $H'$ under other gauge groups varies.

There are several options for the action of $Z_2$ on $SU(3)_c$ and $U(1)_Y$.
The theory with partners of both $SU(3)_c$ and $U(1)_Y$  is nothing but the mirror world (see \cite{Foot:2003eq,Berezhiani:2003xm} for reviews).
The gauge group is $(SU(3)_c\times SU(2)_L\times U(1)_Y )\times (SU(3)_c\times SU(2)_L\times U(1)_Y )'$, and
all of the SM particles have their mirror counterparts.
However, it is not necessary to introduce partners of $SU(3)_c$ and $U(1)_Y$.
Indeed, in the next section, we show that in minimal cases, where $SU(3)_c \times U(1)_Y$ is not replicated, the strong CP problem~\cite{'tHooft:1976up} may be solved by including space-time parity in the $Z_2$ symmetry.

%%%%%%%%%%%%%%%%%%%%%%%%%%
\section{Minimal Models at Scale $v'$}
\label{sec:Models}
%%%%%%%%%%%%%%%%%%%%%%%%%%%

In the previous section, we showed that the essentially unique Higgs structure for breaking $SU(2)_L \times  SU(2)'$ is remarkably minimal, $H(2,1)+H'(1,2)$, implying that in the theory well below scale $v'$ the SM Higgs is $H$. Since the SM Higgs carries non-zero hypercharge and is an $SU(2)'$ singlet, the electroweak group must be extended beyond $SU(2)_L \times  SU(2)'$. Hence, the gauge group with fewest generators that yields our understanding of the SM quartic is $SU(3)_c \times SU(2)_L \times SU(2)' \times U(1)$.  We choose the normalization of the $U(1)$ generator so that it is conventional hypercharge on $SU(2)'$ singlets, so that without loss of generality we have $H(1,2,1, -1/2)$ and $H'(1,1,2, \pm1/2)$.  The vacuum expectation value of $H'$ breaks $SU(2)_R \times U(1) \rightarrow U(1)_Y$.\footnote{In conventional $SU(2)_L \times SU(2)_R$ theories, the breaking of $SU(2)_R$ is accomplished by a triplet.  }

We assume that the $SU(2)_L$ doublet quarks and leptons of the SM are $SU(2)'$ singlets, so they must transform as $q(3, 2,1, 1/6)$ and $\ell(1, 2,1, -1/2)$.  The $Z_2$ symmetry requires that $q,\ell,H$ have partners transforming as $(1,2)$ under $(SU(2)_L,  SU(2)')$.
There are four possible $SU(3)_c\times U(1)$ charge assignments for $(q',\ell',H')$, as listed in Table~\ref{tab:doublets}, where the fermions are left-handed Weyl spinors.  We designate these cases as A$(-,-)$, B$(+,-)$, C$(-,+)$, and D$(+,+)$, where the signs indicates whether $(SU(3)_c,U(1))$ charges are conjugated.   We find that, with minimal field content consistent with gauge anomaly freedom, the right-handed SM quarks and leptons transform as $(1,2)$ under $(SU(2)_L,  SU(2)')$ in model A but as $(1,1)$ in models B, C and D.   In the first sub-section we study Model A
and identify the $Z_2$ partner of $SU(2)_L$ as $SU(2)_R$.  In the second sub-section we study models B, C and D.

%%%%%%%%%
\subsection{$SU(3)_c \times SU(2)_L \times SU(2)_R \times U(1)$}
%%%%%%%%%

\begin{table}[tb]
\begin{center}
\caption{Doublet fields: the four possible $SU(3)_c \times SU(2)_L \times SU(2)' \times U(1)$ assignments for the $Z_2$ partners of 
$q({\bf 3}, 2,1, \frac{1}{6})$, $\ell({\bf 1}, 2,1, -\frac{1}{2})$ and $H({\bf 1}, 2,1, -\frac{1}{2})$.}
\label{tab:doublets}
\vspace{0.2in}
\begin{tabular}{|c|c|c|c|c|}
\hline
	& A$(-,-)$			& B$(+,-)$				& C$(-,+)$		& D$(+,+)$		 \\ \hline
$q'$	& $({\bf \bar{3}}, 1,2, -\frac{1}{6})$	& $({\bf 3}, 1,2, -\frac{1}{6})$	& $({\bf \bar{3}}, 1,2, \frac{1}{6})$	& $({\bf 3}, 1,2, \frac{1}{6})$	 \\
$\ell',H'$	& $({\bf 1}, 1,2, \frac{1}{2})$	& $({\bf 1}, 1,2, \frac{1}{2})$		& $({\bf 1}, 1,2, -\frac{1}{2})$	& $({\bf 1}, 1,2, -\frac{1}{2})$	 \\ \hline
\end{tabular}

\end{center}
\end{table}%

Model A is free of gauge anomalies with $(q,q',\ell,\ell')$.  While there are no gauge-invariant Yukawa couplings, there are interactions between fermions and scalars at dimension 5 
\begin{align}
\label{eq:high_dim}
{\cal L}_A =  \frac{1}{M_u} (q \, \tilde{y}_u q') H^\dag H^{'\dag} +   \frac{1}{M_d} (q \, \tilde{y}_d q') H H' +   \frac{1}{M_e} (\ell \, \tilde{y}_e \ell') H H'  + {\rm h.c.}
\end{align}
Here $\tilde{y}_{u,d,e}$ are dimensionless flavor matrices, with flavor indices suppressed, while $M_{u,d,e}$ are mass scales.

On breaking $SU(2)_R$, the theory below scale $v'$ is the SM (with right-handed neutrinos to be discussed). $q'$ and $\ell'$ contain the $SU(2)_L$-singlet SM quarks and leptons and $U(1)$ is identified as $(B-L)/2$. SM Yukawa couplings arise from (\ref{eq:high_dim}) and are given by $\tilde{y}_{u,d,e} \; v'/M_{u,d,e}$.  

The dimension 5 operators of 
(\ref{eq:high_dim}) can be generated by the exchange of heavy states.  Since the top Yukawa coupling is near unity, at least some of these states must be close to $v'$, and we take these to be fermions, $X$ (and $\bar{X}$ when Dirac), as extra scalars near $v'$ require further fine-tuning.
The possible gauge charges of $X$ for each Yukawa coupling are listed in Table~\ref{tab:X}. 
Anticipating the next section, we also show the possible embedding of these fermions into $SO(10)$ representations with a dimension $210$ or smaller.
In the $f=u,d,e$ sectors, if these heavy fermions have mass matrices $M_{X_f}$ and Yukawa couplings $x_f$ to $q/\ell$ and $x_f'$ to $q'/\ell'$, then the resulting $6 \times 6$ mass matrices are
\begin{align}
\label{eq:6X6}
{\cal M}_f = \begin{pmatrix}
M_{X_f}& x'_f v'\\
x_f v & 0
\end{pmatrix}
\end{align}
where $M_{X_f}, x_f$ and $x_f'$ are $3 \times 3$ matrices. We show the result for $X(1,1)$; for $X(2,2)$ the same result applies except with $v \leftrightarrow v'$ . After integrating out the heavy fermions, we obtain the dimension-5 operator in Eq.~(\ref{eq:high_dim}) with 
\begin{align}
\label{eq:seesaw}
\frac{\tilde{y}_f}{M_f} = x_f \frac{1}{M_{X_f}} x'_f \hspace{0.5in} \mbox{or equivalently} \hspace{0.5in} y_f = x_f \frac{v'}{M_{X_f}} x'_f.
\end{align}
The effect of $Z_2$ on these couplings will be discussed later in this section.

The following dimension-5 operators give masses to both left and right-handed neutrinos,
\begin{align}
\label{eq:high_dimnu}
{\cal L}^\nu =  \frac{1}{M} \left( (\ell \tilde{y} \ell) H^{\dag 2} +  (\ell'  \tilde{y} \ell') H^{'\dag2}\right) + \frac{1}{M_\nu} (\ell \tilde{y}_\nu \ell') H^{\dag}H^{'\dag}
+ {\rm h.c.}
\end{align}
For $M \ll M_\nu |\tilde{y}/\tilde{y}_\nu|(v'/v) $ the left-handed neutrinos obtain Majorana masses, while for $M \gg M_\nu |\tilde{y}/\tilde{y}_\nu|(v'/v)$ they obtain Dirac masses with the right-handed neutrinos in $\ell'$.

\begin{table}[tb]
\caption{Possible $X$ particles for generating Yukawa couplings in Model A.}
\begin{center}
\begin{tabular}{|c||c|c|c|c||c||c||c||}
\hline
              & $SU(3)_c$ &  $SU(2)_L$ &  $SU(2)_R$ &  $U(1)$ & $SU(4)$ & $SO(10)$            & coupling \\ \hline \hline
up          & ${\bf 3}$ & ${\bf 1}$      &${\bf 1}$    & $2/3$ & ${\bf 15}$  & ${\bf 45}$ & $\bar{X} q H^\dag + X q' H^{'\dag}  $ \\
              & ${\bf 3}$ & ${\bf 2}$      &  ${\bf 2}$    & $-1/3$ & ${\bf 6/10}$   & ${\bf 45,54, 210/210}$ &  $\bar{X} q H^{'\dag} + X q' H^\dag$  \\ \hline
down     & ${\bf 3}$ &   ${\bf 1}$    &    ${\bf 1}$  & $-1/3$  & ${\bf 6/10}$  & ${\bf 10,126/120}$ & $\bar{X} q H + X q' H'$ \\
              & ${\bf 3}$ &    ${\bf 2}$  & ${\bf 2}$     & $2/3$  & ${\bf 15}$   & ${\bf  120,126}$ & $\bar{X} q H' + X q' H$ \\ \hline
electron &  ${\bf 1}$ &  ${\bf 1}$   &  ${\bf 1}$   & $-1$  & ${\bf 10}$  & ${\bf  120}$ & $\bar{X} \ell H + X \ell' H' $ \\
              & ${\bf 1}$ &    ${\bf 2}$  & ${\bf 2}$   & $0$ &  ${\bf 1/15}$    & ${\bf  10,120/120, 126}$ & $X \ell H' + X \ell' H$ \\ \hline
neutrino &  ${\bf 1}$ &  ${\bf 1}$   &  ${\bf 1}$   & $0$  & ${\bf 1/15}$  & ${\bf  1,54,210/ 45,210}$ & $X (\ell H^\dagger +  \ell' H'^\dagger) $ \\
              & ${\bf 1}$ &    ${\bf 2}$  & ${\bf 2}$   & $-1$ &  ${\bf 10}$    & ${\bf  210}$ & $ \bar{X} \ell H'^\dagger + X \ell' H^\dagger $ \\
              & ${\bf 1}$ &    ${\bf 3}$  & ${\bf 1}$   & $0$ &  ${\bf 1}$    & ${\bf  45}$ & $ X \ell H^\dagger $ \\
              & ${\bf 1}$ &    ${\bf 1}$  & ${\bf 3}$   & $0$ &  ${\bf 1}$    & ${\bf  45}$ & $ X \ell' H'^\dagger $ \\
              \hline
\end{tabular}
\end{center}
\label{tab:X}
\end{table}%

%%%%%%%%%
\subsection{$SU(3)_c \times SU(2)_L \times SU(2)' \times U(1)$}
%%%%%%%%%
With just $(q,q',\ell,\ell')$, Models B, C and D contain gauge anomalies.
In Models C and D,
$q'$ and $\ell'$ do not have the right charges to be identified with SM $SU(2)_L$-singlet quarks or leptons, and there are no Yukawa-like interactions for electrically-charged fermions at any dimension.
For these theories, the minimal additions for anomaly freedom are  $SU(2)_L$-singlet fermions of the SM, $\bar{u}$, $\bar{d}$ and $\bar{e}$, and their $Z_2$ partners $\bar{u}'$, $\bar{d}'$ and $\bar{e}'$ with gauge charges shown in Table \ref{tab:singlets}. The following Yukawa couplings are allowed for Model C,
\begin{align}
\label{eq:dim4}
{\cal L}_{C} =&  (q \, y_u \bar{u}) H^\dag + (q'  y_u' \bar{u}') H^{'\dag} + (q \, y_d \bar{d}) H + (q'  y_d'  \bar{d}') H'  \nonumber \\
& +  (\ell \, y_e \bar{e}) H +  (\ell'  y_e' \bar{e}')  H' +  (\ell \, \lambda_e \bar{e}') H +  (\ell'  \lambda'_e \bar{e}) H'
+ {\rm h.c.},
\end{align}
where generation indices are suppressed.
In Model D, the allowed Yukawa couplings are
\begin{align}
\label{eq:dim4D}
{\cal L}_{D} =& {\cal L}_C +  (q \, \lambda_u \bar{u}') H^\dag +  (q' \lambda_u' \bar{u}) H^{'\dag} + (q \, \lambda_d \bar{d}') H +  (q' \lambda_d' \bar{d}) H' 
+ {\rm h.c.}.
\end{align}

\begin{table}[tb]
\begin{center}
\caption{Singlet fields: three possible $SU(3)_c \times SU(2)_L \times SU(2)_R \times U(1)$ assignments for the $Z_2$ partners of $\bar{u}({\bf \bar{3}}, 1,1, -\frac{2}{3})$, $\bar{d}({\bf \bar{3}}, 1,1, \frac{1}{3})$, and $\bar{e}({\bf 1}, 1,1, 1)$.} 
\label{tab:singlets}
\vspace{0.2in}
\begin{tabular}{|c|c|c|c|}
\hline
	& B$(+,-)$			& C$(-,+)$				& D$(+,+)$				 \\ \hline
$\bar{u}'$	& $({\bf \bar{3}}, 1,1, \frac{2}{3})$	& $({\bf 3}, 1,1, -\frac{2}{3})$	& $({\bf \bar{3}}, 1,1, -\frac{2}{3})$ \\
$\bar{d}'$	& $({\bf \bar{3}}, 1,1, -\frac{1}{3})$	& $({\bf 3}, 1,1, \frac{1}{3})$	& $({\bf \bar{3}}, 1,1, \frac{1}{3})$ \\
$\bar{e}'$	&	& $({\bf 1}, 1,1, 1)$	& $({\bf 1}, 1,1, 1)$ \\
\hline

\end{tabular}

\end{center}
\end{table}%

After $H'$ obtains a large vacuum expectation value, the partner fermions obtain a large mass and decouple.  The theory has a SM$'$ sector, similar to the SM but at scale $v'$ with weak interactions from $SU(2)'$.  The effective theory below the SM$'$ sector is just the SM (with right-handed neutrinos to be discussed).  In both sectors, the $U(1)$ gauge symmetry is now hypercharge; but in theory C the two sectors have opposite color.

In Model B, $\ell'$ has the right charge to be identified with the SM $SU(2)_L$-singlet lepton, so singlet fields $\bar{e}$ and $\bar{e}'$ are not added. The electron Yukawa couplings is obtained from the third term in Eq.~(\ref{eq:high_dim}). $q'$, on the other hand, cannot be identified with SM $SU(2)_L$-singlet quarks, and hence we add $\bar{u}$, $\bar{d}$, $\bar{u}'$ and $\bar{d}'$ as shown in Table \ref{tab:singlets}, and the up and down Yukawa couplings are as in (\ref{eq:dim4}):
\begin{align}
\label{eq:yukawaB}
{\cal L}_{B} = (q \, y_u \bar{u}) H^\dag + (q'  y_u' \bar{u}') H^{'\dag} + (q \, y_d \bar{d}) H + (q'  y_d'  \bar{d}') H' + \frac{1}{M_e} (\ell \, \tilde{y}_e \ell') H H' 
+ {\rm h.c.}.
\end{align}
 In this hybrid theory, the gauge $U(1)$ can be interpreted as $(B-L)/2$ on leptons and hypercharge on quarks. The SM$'$ sector contains only quarks, and they have opposite hypercharges to the SM quarks.

In Models B and C the $6 \times 6$ mass matrices for the quarks of the two sectors are
\begin{align}
\label{eq:6X6BC}
{\cal M}_f = \begin{pmatrix}
y'_f v'& 0\\
0 & y_f v
\end{pmatrix},
\end{align}
while in Model D they are
\begin{align}
\label{eq:6X6D}
{\cal M}_f = \begin{pmatrix}
y'_f v'& \lambda_f' v'\\
\lambda_f v  & y_f v
\end{pmatrix}.
\end{align}

Neutrino masses are generated by the operators of (\ref{eq:high_dimnu}), as in Model A.

Models B, C and D contain heavy exotic matter. The exotic leptons (present only in Models C and D) mix with the SM leptons via (\ref{eq:dim4}) and are unstable.  The exotic quarks of Model D mix with the SM quarks via (\ref{eq:dim4D}) and are unstable.
However, in Models B and C, exotic quarks cannot mix with SM quarks, so that the lightest exotic quark, $u'$, is stable.

%%%%%%%%%
\subsection{$SU(4)\times SU(2)_L \times SU(2)_R$}
%%%%%%%%%
In Models A and D, we can embed $SU(3)_c\times U(1)$ into the Pati-Salam group $SU(4)$ \cite{Pati:1974yy}. $H$ and $H'$ are embedded into $({\bf 4},{\bf 2},{\bf 1})$ and $({\bf 4},{\bf 1},{\bf 2})$, respectively. The vev of $H'$ breaks the $SU(4)\times SU(2)_L \times SU(2)'$ symmetry to the SM gauge group. The stability of this vacuum, as well as possible quantum corrections to the SM Higgs quartic coupling, is discussed in Appendix~\ref{sec:422 at vR}.

The embedding of $(q,q',\ell,\ell',H,H')$ is given in Table~\ref{tab:doublets422}.
Again, Model D contains gauge anomalies and $(q',\ell')$ do not have the right charges to be identified with SM $SU(2)_L$-singlet quarks or leptons: further $SU(2)_L$-singlet fermions must be added. Their $SU(4)\times SU(2)_L \times SU(2)'$ charges are given by embedding the particles in Table~\ref{tab:singlets} into $SU(4)\times SU(2)_L \times SU(2)'$.
Here we focus on Model A where $q'$ are the SM $SU(2)_L$-singlet quarks, and we identify $SU(2)'$ with $SU(2)_R$.

Interactions between fermions and Higgs occur via dimension 5 operators. Up, down and electron masses arise from
\begin{align}
\label{eq:high_dim422}
{\cal L}_{422_A} =  \frac{1}{M_u} \, ( \psi^a \tilde{y}_u  \psi'_{[a}) \, H^{'\dag b}H^\dag_{b]}  +  \frac{1}{M_d} \, ( \psi^{[a} \tilde{y}_d \psi'_{a})\, H^{b]} H'_ b +  \frac{1}{M_e} \, (\psi^a \tilde{y}_e \, \psi'_{b}) \,H'_ a H^b  +  {\rm h.c.}
\end{align}
where $a,b$ are $SU(4)$ indices.  Note the antisymmetrization in the first two terms, corresponding to the exchange of a fermionic {\bf 6}. Inserting $H^4=H$ and $H'_4 = H'$ yields (\ref{eq:high_dim}).  These operators result from integrating out heavy Dirac fermions $X$ in the $(6,2,2), (6,1,1)$ and $(1,2,2)$ representations, respectively.  Similarly, neutrino masses arise from
\begin{align}
\label{eq:high_dim422nu}
{\cal L}^\nu_{422_A} =  \frac{1}{M}  [\, (\psi^a \tilde{y} \, \psi^b) \, H^\dag_a H^{\dag}_b +  \, (\psi'_{a} \tilde{y} \, \psi'_{b}) \, H'^{\dag a} H'^{\dag b}] +  \frac{1}{M_\nu}  \, (\psi^a \tilde{y}_\nu \, \psi'_{b}) \, H^\dag_a H^{'\dag b}  +{\rm h.c.}.
\end{align}
which yield the operators of (\ref{eq:high_dimnu}). 

\begin{table}[tb]
\caption{$SU(4)\times SU(2)_L \times SU(2)_R$ embedding of $SU(2)$ doublets.}
\begin{center}
\begin{tabular}{|c|c|c|c|c|}
\hline
	& $(q,\ell)\equiv \psi$		&$(q',\ell')\equiv \psi'$& $H$				& $H'$ \\ \hline
A	& $({\bf 4},{\bf 2},{\bf 1})$	& $({\bf \bar{4}},{\bf 1},{\bf 2})$	& $({\bf 4},{\bf 2},{\bf 1})$	& $({\bf \bar{4}},{\bf 1},{\bf 2})$ \\ \hline
D	& $({\bf 4},{\bf 2},{\bf 1})$	& $({\bf 4},{\bf 1},{\bf 2})$		& $({\bf 4},{\bf 2},{\bf 1})$	& $({\bf 4},{\bf 1},{\bf 2})$ \\ \hline
\end{tabular}
\end{center}
\label{tab:doublets422}
\end{table}%

%%%%%%%%%%%%%%%%%%%%%%%%%%
\section{$Z_2$ Symmetry and the Strong CP Problem}
\label{sec:CP}
%%%%%%%%%%%%%%%%%%%%%%%%%%%

If the action of the $Z_2$ symmetry is the simple exchange of fermions, $\psi \leftrightarrow \psi'$, the Yukawa couplings are required to satisfy
$\tilde{y}_{ij}= \tilde{y}_{ji}$ for Eqs.~(\ref{eq:high_dim}, \ref{eq:high_dim422}), or $y_{ij}'=y_{ij}$ as well as $\lambda_{ij}'=\lambda_{ij}$   for Eqs.~(\ref{eq:dim4},\ref{eq:dim4D}).
If, on the other hand, the $Z_2$ symmetry involves the space-time parity transformation, $\psi(t,x) \rightarrow i \sigma^2 \psi^{'*}(t,-x)$, the requirements are $\tilde{y}_{ij}= \tilde{y}_{ji}^{*}$ (i.e.~Hermitian) or $y_{ij}' = y_{ij}^*$ and $\lambda_{ij}' = \lambda_{ij}^*$ . Here $\sigma^2$ acts on the spinor index.
The contribution to the strong CP phase from the quark Yukawa couplings is proportional to the phases of
\begin{align}
{\rm det} \tilde{y},~
{\rm det}y \times {\rm det}y^*,~
{\rm det}
\begin{pmatrix}
y^* & ~\lambda^* \\
\lambda & y
\end{pmatrix},
\end{align}
for Models A, B and C, and D, which vanish.
Space-time parity also forces $\theta_{QCD}=0$, so that $\bar{\theta}=0$ at tree-level.

We summarize this solution of the strong CP problem by
\begin{align}
\label{eq:strongCP}
P: \hspace{0.25in} \psi(t,x) \rightarrow i \sigma^2 \psi^{'*}(t,-x) \implies 
 \begin{tabular}{c}
 \hspace{-0.6in} $ \tilde{y}^\dagger = \tilde{y}$ \hspace{0.9in} \mbox{Model A in (\ref{eq:high_dim}, \ref{eq:high_dim422})} \\
 $y' = y^*$ and $\lambda' = \lambda^*$  \mbox{Models B, C, D in (\ref{eq:dim4},\ref{eq:dim4D})}\\
 \hspace{-2.7in} $\theta_{QCD} = 0$
  \end{tabular}
\end{align}
which applies whether the gauge group is 3221 or 422.  $P$ does not forbid phases in $\tilde{y}$ or $y$, and the CKM phase is obtained as usual from the physical phase of $\tilde{y}$ or $y$.
For cases where $SU(2)'$ is identified as $SU(2)_R$, we call the parity transformation $P_{LR}$, since it exchanges $SU(2)_L \leftrightarrow SU(2)_R$.  

In Model A, before the heavy Dirac $X$ fermions are integrated out, the $6 \times 6$ color triplet mass matrices take the form of (\ref{eq:6X6}).  At this level it is easy to see that the strong CP problem is solved because $P_{LR}$ forces $M_{X_{u,d}}^\dagger = M_{X_{u,d}}$ and $x'_{u,d} = x^\dagger_{u,d}$.  For models B and C the $6 \times 6$ quark mass matrices take the form of (\ref{eq:6X6BC}). Since $y'_{u,d} = y_{u,d}^*$, the contribution to $\bar{\theta}$ from the SM$'$ sector cancels that from the SM sector.

Model A, with gauge group 3221 and $\bar{X}$ having the same charge as $\bar{u}$, $\bar{d}$ and $\bar{e}$, was proposed in Ref.~\cite{Babu:1989rb} as a solution to the strong CP problem based on parity. There parity was assumed to be softly broken to obtain the hierarchy between the vacuum expectation values of $H'$ and the electroweak scale. As we have shown in the previous section, soft breaking is not required. Parity symmetry can be spontaneously broken by the condensation of $H'$, thereby explaining the vanishing of the SM quartic at scale $v'$.

Since parity is spontaneously broken, the strong CP phase may be generated by higher-dimensional operators. The following operator is composed only of bosonic fields, and is not controlled by any symmetry at scale $v'$,%
\footnote{The operator may be controlled by a symmetry realized at a high energy scale. For example, in the $SO(10)$ model discussed in the next section, the symmetry at high scales is $SO(10) \times CP$ and CP forbids the operator in Eq.~(\ref{eq:HHGG}). After $SO(10)\times CP$ symmetry breaking, the operator is generated with a further suppression factor of $v_G/ M_*$. Supersymmetry can also suppress the operator.}
\begin{align}
\label{eq:HHGG}
{\cal L}_6 = \frac{1}{M_*^2} (|H^2| - |H'|^2) \, G \tilde{G},
\end{align}
where $M_*$ is a cut-off scale and $G$ is the field strength of $SU(3)_c$. Condensation of $H'$ yields the strong CP phase
\begin{align}
\theta \sim 8 \pi^2 \left(\frac{v'}{M_*} \right)^2 \sim 10^{-9} \left( \frac{v'}{10^{13}~{\rm GeV}} \right)^2 \left( \frac{2.4 \times 10^{18}~{\rm GeV}}{M_*} \right)^2.
\end{align}
For a cut-off scale of the Planck mass, satisfying the experimental constraint $\theta < 10^{-10}$~\cite{Crewther:1979pi,Baker:2006ts,Graner:2016ses} requires $v' < 10^{13}$ GeV. 

Quantum corrections also gives a non-zero strong CP phase. In Models B and C they are essentially the same as in the SM model, and are negligibly small~\cite{Ellis:1978hq}.
The quantum correction in Model D is not known and will be investigated elsewhere.
In Model A, after integrating out heavy states $X \bar{X}$, any radiative corrections to the dimension 5 operators do not induce $\bar{\theta}$; but contributions arise from parity-invariant operators of dimension 7
\begin{align}
{\cal L}_{\rm eff} = \frac{1}{M_u^3} \, q \left( c_u |H|^2 + c_u^\dag |H'|^2 \right) q' \; H^\dag H^{'\dag} + \frac{1}{M_d^3} \, q \left( c_d |H|^2 + c_d^\dag |H'|^2 \right) q' \; H H' + {\rm h.c.}.
\end{align}
After $H'$ obtains a vev, this leads to non-Hermitian Yukawa couplings. As shown in Appendix \ref{sec:correction}, non-Hermitian contributions to the flavor matrices $c_u$ and $c_d$ arise first at 2-loop level, and require flavor mixing.%
\footnote{Ref.~\cite{Babu:1989rb} shows that the one-loop correction to $\theta$ is absent. It also claims that the two-loop corrections are suppressed by $v /v'$, which we could not confirm.}
The typical correction is of the form
\begin{align}
\label{eq:typcorr}
\Delta \theta \sim \frac{g^4}{(16\pi^2)^2} |V_{cb}|^2 \, C \simeq 6\times 10^{-9} \, C,
\end{align}
where $g$ is a gauge coupling constant at scale $v'$ and $C$ is a numerical constant which depends on the theory of flavor. The constant $C$ is {\it at most} $O(10^3)$ and such large values result when there is no hierarchy among the various $X$ masses. However, with a Frogatt-Nielsen structure the quark mass hierarchy naturally follows from an $X$ mass hierarchy, and this gives small values for $C$, with the leading contribution typically
\begin{align}
C \sim \frac{y_b^3}{y_s} \sim 3\times10^{-3}.
\end{align}
Note that $|V_{cb}|^2$ of (\ref{eq:typcorr}) arises from the product of the 23 mixing in the up and down sectors, $\theta^u_{23}\theta^d_{23}$, so that $\Delta \theta$ is further suppressed if one is much smaller than the other.

%%%%%%%%%
\section{Gauge Coupling Unification}
\label{sec:gcu}
%%%%%%%%%

We investigate the running of gauge couplings in the LR theories to determine whether a more precise unification is possible than in the SM, which requires large threshold corrections to avoid exclusion from proton decay.  We anticipate Section \ref{sec:so10} where we show that these theories can be successfully embedded into $SO(10)$.  We treat separately the cases where the gauge group above $v'$ is $SU(3)_c\times SU(2)_L \times SU(2)_R \times U(1)_{B-L}$ and $SU(4)\times SU(2)_L \times SU(2)_R$.

\subsection{$SU(3)_c\times SU(2)_L \times SU(2)_R \times U(1)_{B-L}$ }

The gauge couplings evolve from IR to UV as follows.
Between the electroweak scale and the scale $v'$, $SU(3)_c \times SU(2)_L \times U(1)_Y$ couplings $(g_3, g_2, g_1)$ evolve as in the SM,
\begin{align}
\label{eq:rge_sm}
\frac{{\rm d}}{{\rm dln} \mu} \left( \frac{8\pi^2}{g_i^2(\mu)} \right)
= b_i,~~~~~~~(b_3,b_2,b_1) = (7,\frac{19}{6},-\frac{41}{10}),
\end{align}
where the $U(1)_Y$ coupling $g_1$ is suitably normalized for unification, $g_1^2 = 5/3 g^{'2}$.
At the scale $v'$, the $U(1)$ coupling $g_{B-L}$, suitably normalized for unification, is obtained from the relation
\begin{align}
\frac{1}{g_1^2(v')} = \frac{2}{5} \frac{1}{g_{B-L}^2(v')} +  \frac{3}{5}\frac{1}{g_2^2(v')}.
\end{align}
The $SU(2)_R$ coupling is the same as the $SU(2)_L$ coupling. 

Above the scale $v'$, $SU(3)_c\times SU(2)_L \times SU(2)_R \times U(1)_{B-L}$ couplings evolve towards the unification scale $M_G$,
\begin{align}
\frac{{\rm d}}{{\rm dln} \mu} \left( \frac{8\pi^2}{g_i^2(\mu)} \right) = b_i,~~~~~~~(b_3,b_2,b_{B-L}) = (7,\frac{19}{6},-\frac{9}{2})  + (\Delta b_3, \Delta b_{2},\Delta b_{B-L} ),
\end{align}
where $\Delta b_i$ denotes the contribution from heavy $X$ states. We assume that the $X$ states form nearly degenerate $SO(10)$ multiplets, so that $\Delta b_i$ do not affect relative running of the gauge couplings.  If there are many large $X$ multiplets that are light, the unified gauge coupling grows so that unified threshold corrections and two loop running effects are expected to give contributions to $\Delta(M_G)$, defined below in (\ref{eq:Delta}), in excess of 10.  However, with $X$ in 10 and 45 dimensional representations this is easy to avoid by taking
\begin{align}
\prod_a \left( \frac{M_{10_a}}{M_G} \right)^{1/8} \prod_b \left( \frac{M_{45_b}}{M_G} \right) > 10^{-23} \left(\frac{10^{16}~{\rm GeV}}{M_G}\right)^{21/16},
\end{align}
where generation indices $a,b$ run over multiplets $X_a$ lighter than $M_G$.

To quantify the quality of the unification, we define
\begin{align}
\label{eq:Delta}
\frac{8\pi^2}{ \bar{g}(\mu)^2} \equiv & \frac{8\pi^2}{3} \left( \frac{1}{g_{\rm B-L}^2 (\mu)} + \frac{1}{g_{2}^2 (\mu)} +\frac{1}{g_{3}^2 (\mu)}   \right), \nonumber \\
\Delta^2(\mu) = & \frac{1}{3}\sum_i \left( \frac{8\pi^2}{ \bar{g}(\mu)^2} - \frac{8\pi^2}{ g_{i}(\mu)^2}   \right)^2, \nonumber \\
\Delta_{\rm min} = & \;{\rm min}_{\mu} \Delta (\mu).
\end{align}

\begin{figure}[htbp]
\centering
\includegraphics[width=0.7\linewidth]{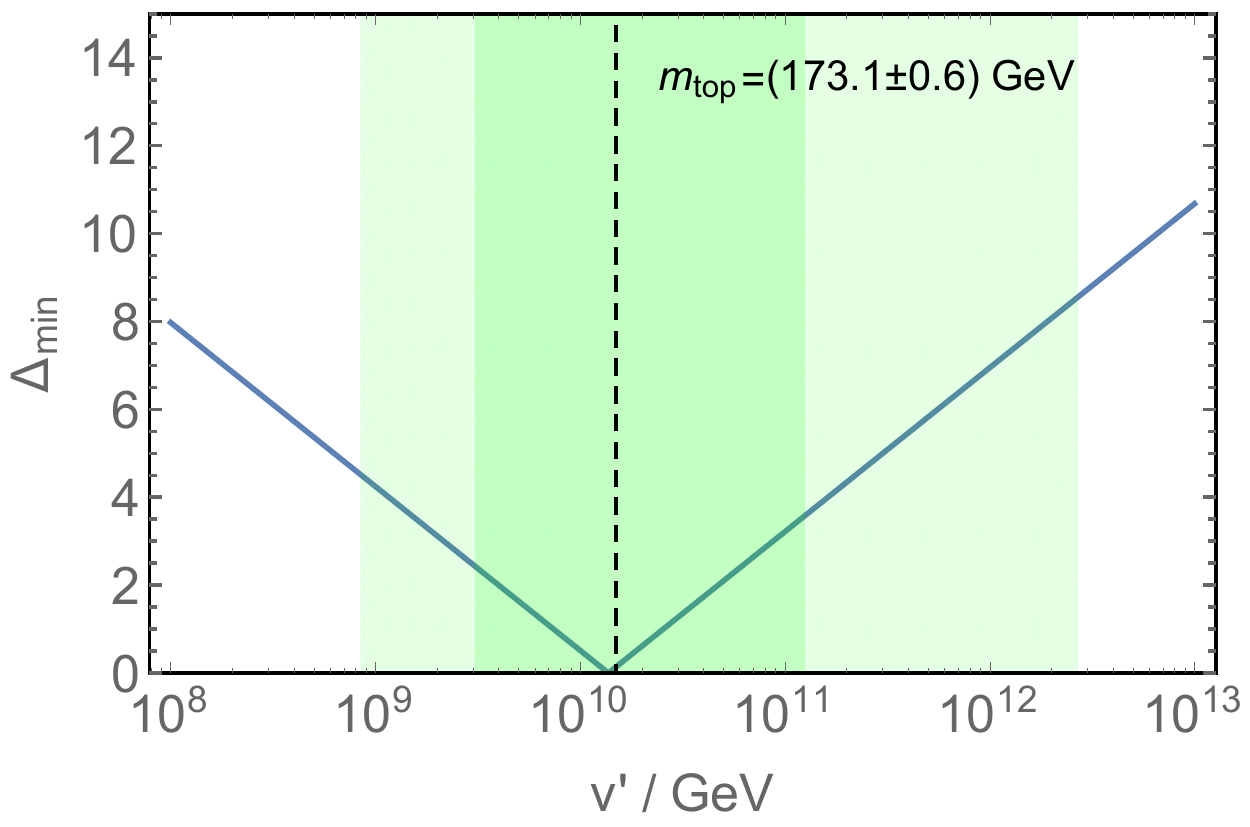}
\caption{\textsl{\small
The quality of gauge coupling unification as a function of $v'$. The vertical dark (light) band shows the prediction for $v'$ from the vanishing Higgs quartic coupling at $1\sigma$ ($2\sigma$).
}}
\label{fig:unification}
\end{figure}

In Fig.~\ref{fig:unification}, the solid curve shows the value of $\Delta_{\rm min}$ as a function of the scale $v'$. The shaded band shows the scale at which the quartic coupling of the SM Higgs vanishes for the top mass of $173.1\pm 0.6$ GeV, with the vertical dashed line showing the central value.  Interestingly, around this central value precise gauge coupling unification is achieved.  Note however that the heavy states from each $SO(10)$ multiplet (e.g.~the gauge bosons, the $SO(10)$ symmetry breaking field, and $X\bar{X}$) are not expected to be degenerate and will typically have mass ratios of $O(1)$. This is expected to generate a threshold correction to $\Delta$ of $\sim 10$. Similar contributions may arise from 2-loop running.  Thus the remarkable agreement between gauge coupling unification and the observed value of the Higgs boson mass, allows for $v'$ anywhere in the range of $(10^9 - 3 \times 10^{12})$ GeV at the $2\sigma$ level.  A more precise determination results if the uncertainties on the top quark mass are reduced.

\begin{figure}[htbp]
\centering
\includegraphics[width=0.7\linewidth]{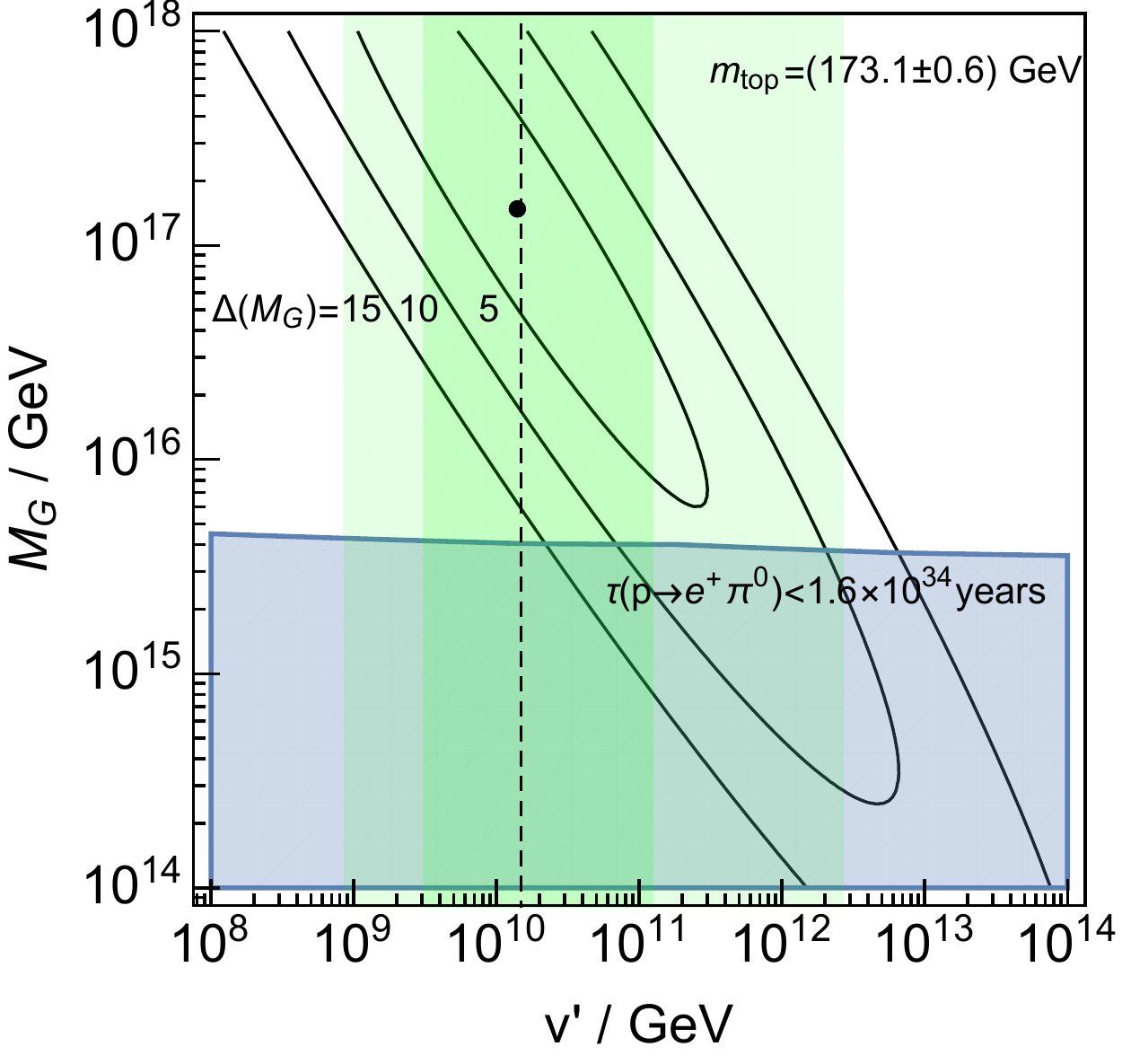}
\caption{\textsl{\small
The quality of gauge coupling unification as a function of $v'$ and the unification scale $M_G$. The vertical band shows the prediction for $v'$ from the vanishing Higgs quartic coupling.
}}
\label{fig:unification2}
\end{figure}

Fig.~\ref{fig:unification2} shows contours of $\Delta (M_G)$ in the $(v', M_G)$ plane, and the constraint on $M_G$ from the proton decay~\cite{Miura:2016krn}. The parameter point which minimizes $\Delta$ has a large $M_G$, and thus cannot be probed by near future searches for nucleon decay. However, the above mentioned threshold corrections to $\Delta$ of $\sim 10$ implies that there is an interesting region of parameter space with lower $M_G$ that will be probed by near future searches.

%%%%%%%%%%%%%%%%%%%%%%%%%%%%%%%%%%%%%%%%%%%%%%%%%%%%%%%%%%%%%
%%%%%%%%%%%%%%%%%%%%%%%%%%%%%%%%%%%%%%%%%%%%%%%%%%%%%%%%%%%%%%%%%%%%%%%%

%%%%%%%%%%%%%%%%%%%%%%%%%%%%%%%%%%%%%%%%%%%%%%%%%%%%%%%%%%%%%
%%%%%%%%%%%%%%%%%%%%%%%%%%%%%%%%%%%%%%%%%%%%%%%%%%%%%%%%%%%%%%%%%%%%%%%%

\subsection{$SU(4)\times SU(2)_L \times SU(2)_R$ }

The gauge couplings evolve from IR to UV as follows. Between the electroweak scale and the scale $v'$, $SU(3)_c \times SU(2)_L \times U(1)_Y$ couplings $(g_3, g_2, g_1)$ evolve as in the SM, Eq.~(\ref{eq:rge_sm}). At the scale $ v'$, $U(1)_Y$ and $SU(3)_c$ are embedded into $SU(4) \times SU(2)_R$, with the relation between gauge couplings given by
\begin{align}
\frac{1}{g_1^2 (v')} =  \frac{2}{5} \frac{1}{g_3^2 (v')} + \frac{3}{5} \frac{1}{g_2^2 (v')}.
\end{align}
This fixes $v'$ to be around $6 \times 10^{13}$ GeV.
Above the scale $v'$, $SU(4)\times SU(2)_L \times SU(2)_R$ couplings evolve toward the unification scale as
\begin{align}
\frac{{\rm d}}{{\rm dln} \mu} \left( \frac{8\pi^2}{g_i^2(\mu)} \right) = b_i,~~~~~~~(b_4,b_2) = (\frac{28}{3},2)  + (\Delta b_4, \Delta b_{2}).
\end{align}
The $SU(4)\times SU(2)_L \times SU(2)_R$ gauge couplings meet at a scale near $4 \times 10^{15}$ GeV.

\begin{figure}[htbp]
\centering
\includegraphics[width=0.7\linewidth]{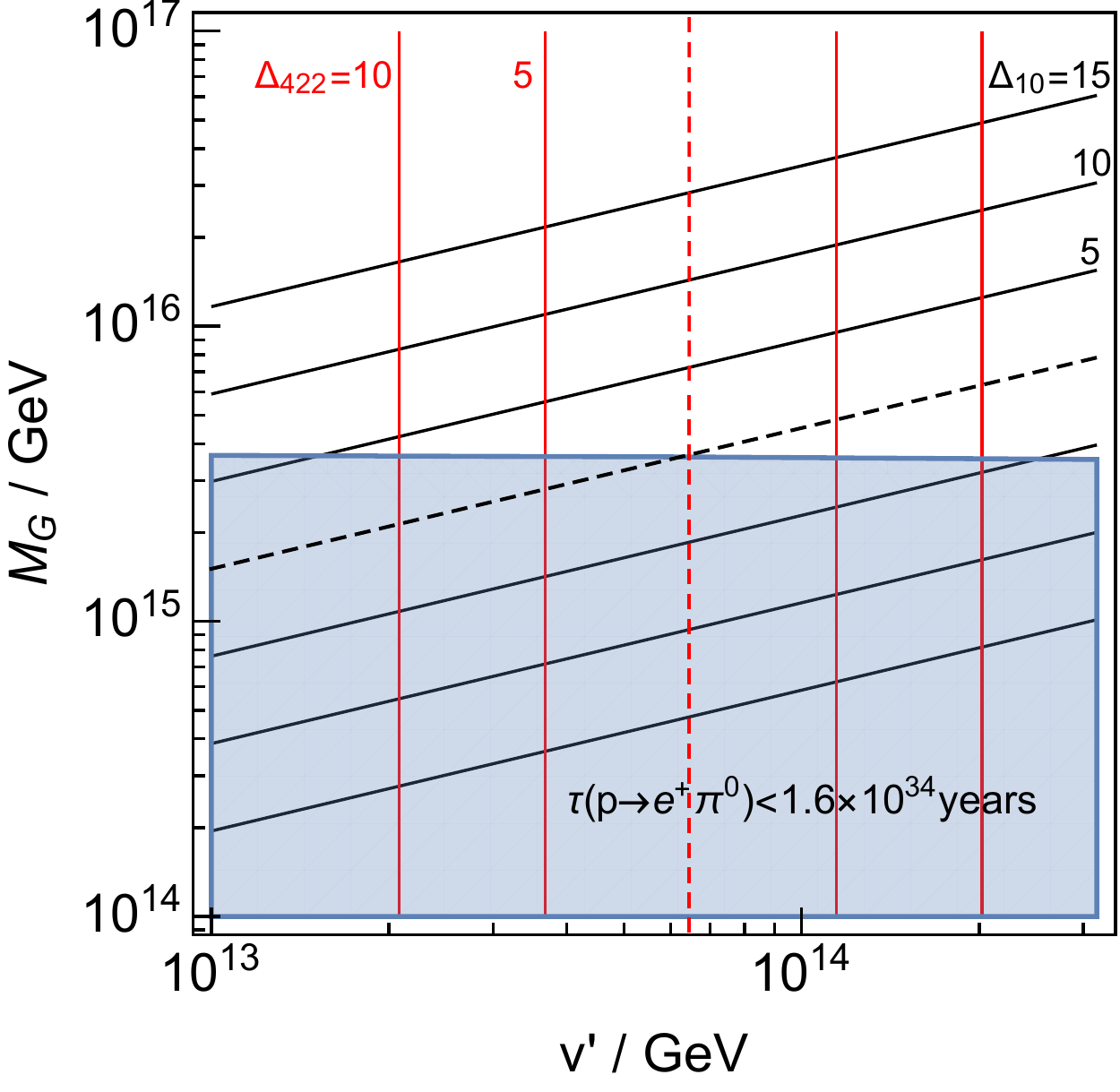}
\caption{\textsl{\small
The quality of gauge coupling unification as a function of $v'$ and the unification scale $M_G$.
}}
\label{fig:unification422}
\end{figure}

We quantify the quality of the unification in the following way. For a given $v'$, we define
\begin{align}
\Delta_{422}(v') \equiv   \left| \frac{3}{5} \frac{8\pi^2}{g_2^2 (v')} + \frac{2}{5} \frac{8\pi^2}{g_3^2 (v')} - \frac{8\pi^2}{g_1^2 (v')} \right|
\end{align}
We then match the gauge couplings at $v'$, $g_4(v') = g_3 (v')$, and evolve them towards the unification scale $M_G$, where we define
\begin{align}
\Delta_{10}(M_G) \equiv \left| \frac{8\pi^2}{g_4^2 (M_G)} - \frac{8\pi^2}{g_2^2 (M_G)} \right|.
\end{align}
In Fig.~\ref{fig:unification422}, we show contours of $\Delta_{422}$ and $\Delta_{10}$ in the $(v', M_G)$ plane.
In the parameter region with $\Delta_{422},\Delta_{10} \lsim 10$, proton decay may be observed in near future experiments.

The embedding of $SU(3)_c \times U(1)_Y$ into $SU(4) \times SU(2)_R$ requires that the scale $v'$ is around $6 \times 10^{13}$ GeV.
If $\lambda_{\rm SM}(v')=0$, this requires a top quark mass below the observed central value.
However, quantum corrections from the colored Higgses give $\lambda_{\rm SM}(v')<0$ (see Appendix~\ref{sec:422 at vR}).  While the size of these corrections are typically of order $O(10^{-3})$, there are parameter regions with $\lambda_{\rm SM}(v') \sim -10^{-2}$, allowing top quark masses close to the observed central value.

%%%%%%%%%%%%%%%%%%%%%%%%%%%%%%%%%%%%%%%%%%%%%%%%%%%%%%%%%%%%%
%%%%%%%%%%%%%%%%%%%%%%%%%%%%%%%%%%%%%%%%%%%%%%%%%%%%%%%%%%%%%%%%%%%%%%%%

%%%%%%%%%%%%%%%%%%%%%%%%%%
\section{$SO(10)$ Unification}
\label{sec:so10}
%%%%%%%%%%%%%%%%%%%%%%%%%%%

In this section we embed Model A into $SO(10)$ grand unified theories.
In Model A, $q'$ and $\ell'$ have the opposite $SU(3)_c \times U(1)$ charges to $q$ and $\ell$. Thus in this section we denote them as $\bar{q}'$ and $\bar{\ell}'$.

\subsection{Matter unification}

The embedding of $q$, $\ell$, $H$ and their parity partners into $SO(10)$ multiplets is shown in Table~\ref{tab:so10}. We also show the embedding into the $SU(4)\times SU(2)_L \times SU(2)_R$ subgroup. Quarks and leptons are unified into the ${\bf 16}$ of $SO(10)$;
the $U(1)$ gauge symmetry is nothing but $B-L$.
The Higgs doublets $H$ and $H'$ are also unified into a ${\bf 16}$ or ${\bf 144}$; they cannot live in a ${\bf 10}$, since this contains a $(1,2,2)$.
The embedding of the heavy states $X\bar{X}$ is shown in Table~\ref{tab:X}.
Introduction of $X \bar{X}$ with a large representation leads to a large gauge coupling at the unification. Thus we focus on the cases where $X \bar{X}$ are ${\bf 10}$, ${\bf 45}$ or ${\bf 54}$.

\begin{table}[htp]
\caption{$SO(10)$ embedding of quarks, leptons and Higgses.}
\begin{center}
\begin{tabular}{|c|c|c|c|c|}
\hline
								& $(q,\ell)$				& $(\bar{q}',\bar{\ell}')$		& $H$				& $H'$ \\ \hline
$SU(4)\times SU(2)_L \times SU(2)_R$	& $({\bf 4},{\bf 2},{\bf 1})$	& $({\bf \bar{4}},{\bf 1},{\bf 2})$	& $({\bf 4},{\bf 2},{\bf 1})$	& $({\bf \bar{4}},{\bf 1},{\bf 2})$ \\ \hline
$SO(10)$							& \multicolumn{2}{c|}{${\bf 16} \equiv \psi_{16}$}		& \multicolumn{2}{c|}{${\bf 16},{\bf 144} \equiv \phi_{H}$}		\\ \hline
\end{tabular}
\end{center}
\label{tab:so10}
\end{table}%

%%%%%%%%
\subsection{$SO(10)$ symmetry breaking and $Z_2$ symmetry}
%%%%%%%%

%%%%%%%%%%%%%%%%%%%%%%%%%%%%%%%%%%%%%%%%%%%%%%%%%%%%%%%%%%%%%
\begin{figure}[tb]
\centering
\includegraphics[width=0.7\linewidth]{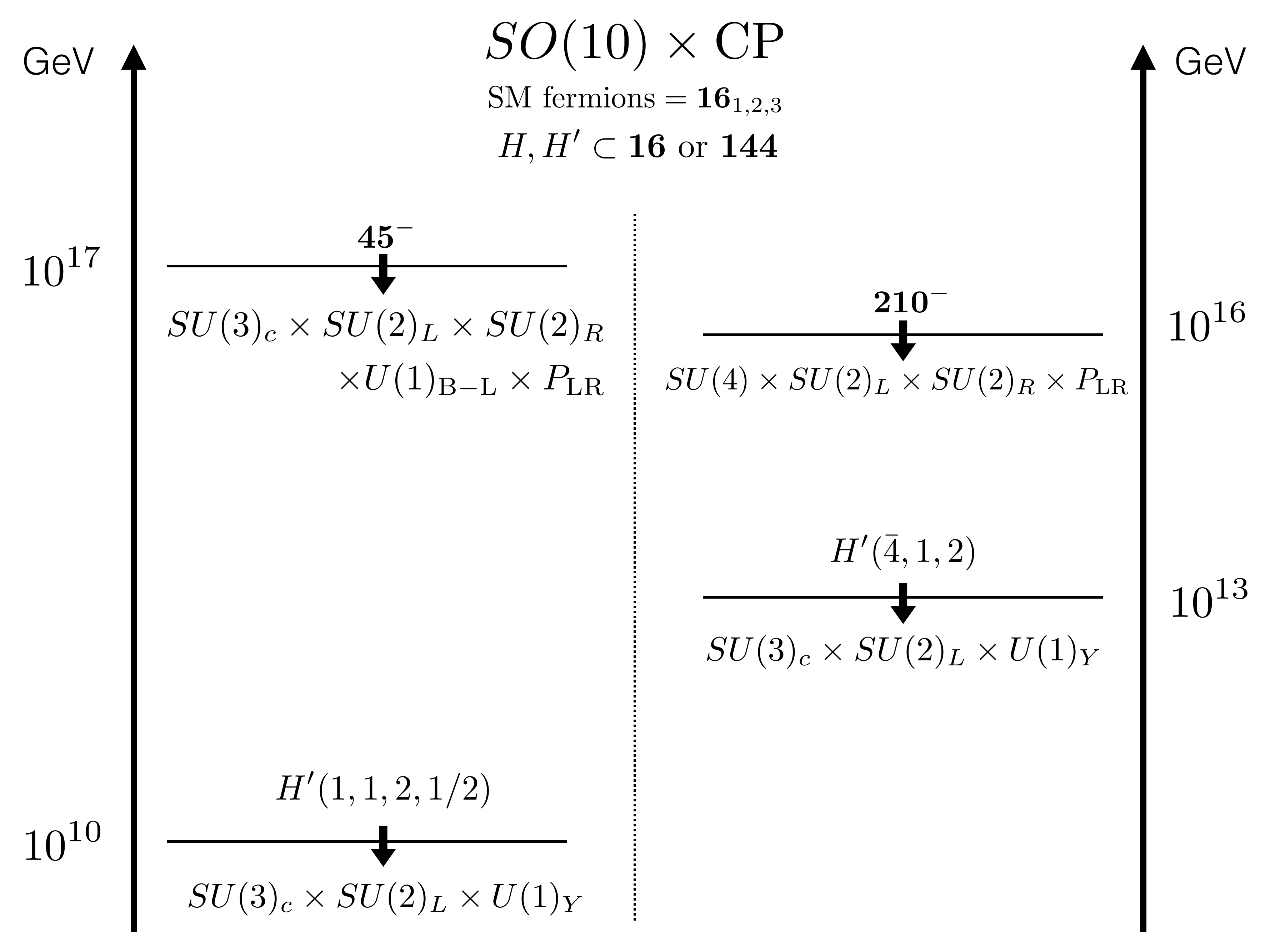}
\includegraphics[width=0.7\linewidth]{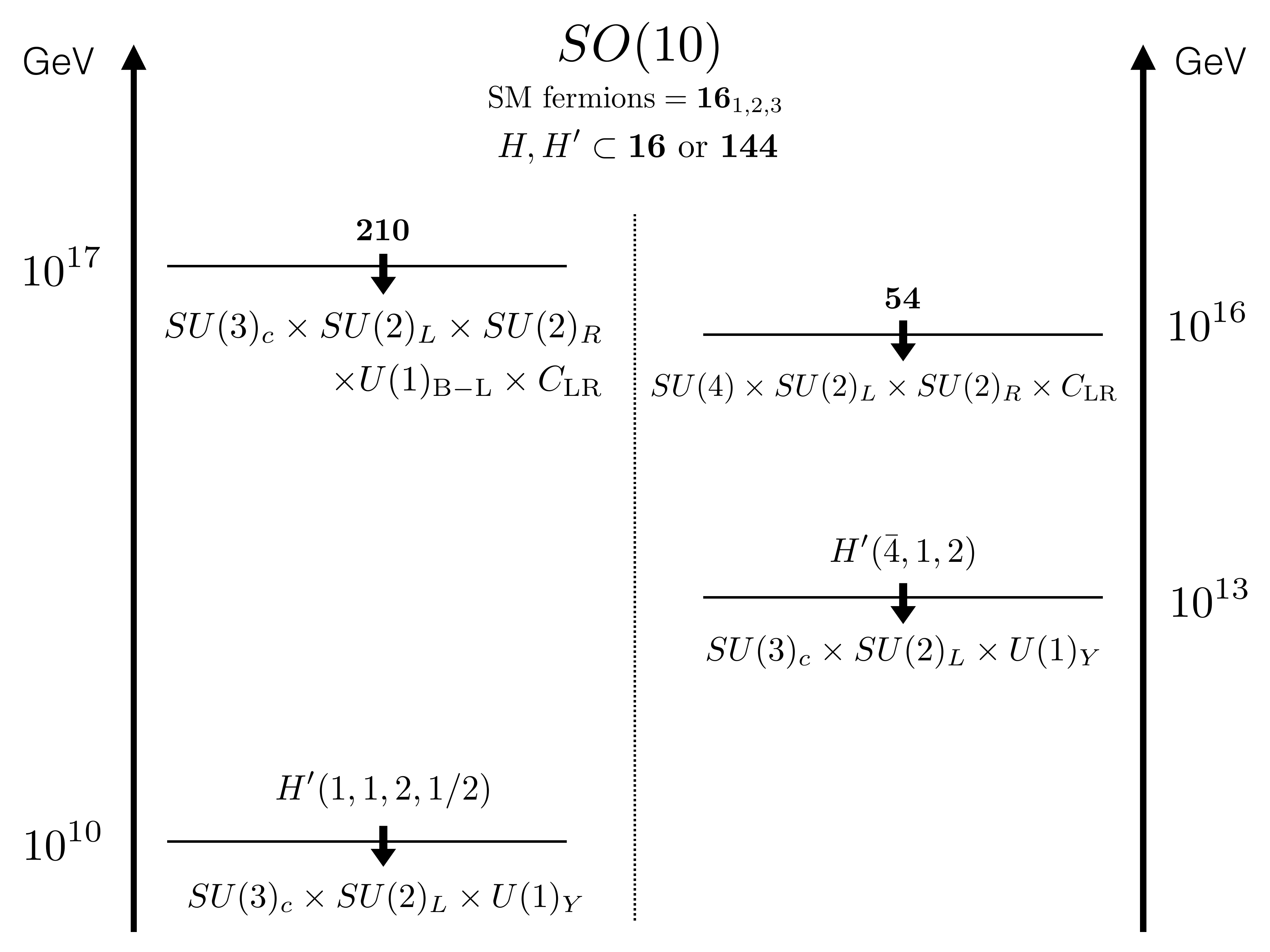}
\caption{\textsl{\small
The symmetry breaking of $SO(10)\times CP$ to the SM gauge group that solves the strong CP problem and explains the vanishing quartic coupling (top).
The symmetry breaking of $SO(10)$ to the SM gauge group can also explain the vanishing quartic coupling (bottom).
}}
\label{fig:chain}
\end{figure}
%%%%%%%%%%%%%%%%%%%%%%%%%%%%%%%%%%%%%%%%%%%%%%%%%%%%%%%%%%%%%%%%%%%%%%%%

In the previous section we consider two types of $Z_2$ symmetries. One is the simple exchange of fields, $(\ell,q)\rightarrow (\bar{\ell}',\bar{q}')$. We call this $Z_2$ transformation $C_{\rm LR}$, as it involves charge conjugation and the exchange of $SU(2)_L$ with $SU(2)_R$. The other $Z_2$ involves space-time parity, with $(t,x) \rightarrow (t,-x)$ and $(\ell,q) \rightarrow i \sigma^2 \; (\bar{\ell}',\bar{q}')^{*}$, which we call $P_{\rm LR}$.

$C_{\rm LR}$ is present in all $SO(10)$ theories, as it is a subgroup of the $SO(10)$ gauge group, and is called $D$ parity in the literature~\cite{Chang:1983fu,Chang:1984uy}.  On the other hand, as in the SM, space-time parity is not a symmetry of $SO(10)$ theories because the fermions are chiral.  Nevertheless, if the $SO(10)$ theory is CP invariant it is also $P_{\rm LR}$ invariant, because
\begin{align}
P_{\rm LR} = CP * C_{\rm LR}.
\end{align}
For some patterns of the $SO(10)$ symmetry breaking, 
$C_{\rm LR} (P_{\rm LR})$ remains unbroken;
the vanishing Higgs quartic coupling is then explained by $SO(10)$ gauge symmetry (and $CP$).
Fig.~\ref{fig:chain} shows the required symmetry breaking pattern.
A scalar ${\bf 54}$ ($\equiv \phi_{54}$) and ${\bf 210}$ ($\equiv \phi_{210}$)  give the symmetry breaking patterns
\begin{align}
\label{eq:54SB}
SO(10) \;\;  \stackrel{\phi_{54}}{\longrightarrow} \;\;  SU(4) \times SU(2)_L \times SU(2)_R \times C_{LR},
\end{align}
\begin{align}
\label{eq:210SB3221}
SO(10) \;\;  \stackrel{\phi_{210}}{\longrightarrow} \;\;  SU(3) \times SU(2)_L \times SU(2)_R \times U(1)_{B-L} \times C_{LR},
\end{align}
respectively.

Having $P_{\rm LR}$ unbroken below the unified scale is particularly interesting, as it solves the strong CP problem.
This requires the symmetry above the unification scale to be $SO(10) \times CP$.  $P_{\rm LR}$ remains unbroken if the $SO(10)$ symmetry breaking field vev is odd under both $C_{\rm LR}$ and $CP$.   The smallest representation that leaves $SU(2)_L \times SU(2)_R$ unbroken via a vev odd under $C_{\rm LR}$ is ${\bf 45}$ ($\equiv \phi_{45}$),~\cite{Chang:1983fu,Chang:1984uy}, giving the symmetry breaking pattern%
\footnote{The $SO(10)$ symmetry breaking into $SU(3) \times SU(2)_L \times SU(2)_R \times U(1)_{B-L}$ solely by the vev of ${\bf 45}$ is not possible at the tree level because of tachyonic directions~\cite{Yasue:1980fy,Yasue:1980qj,Anastaze:1983zk}. It is required to add extra GUT symmetry breaking fields e.g.~${\bf 54}$~\cite{Babu:1984mz} or lift the tachyonic directions by quantum corrections~\cite{Bertolini:2009es}.}
\begin{align}
\label{eq:45SB}
SO(10) \times CP \;\;  \stackrel{ \phi_{45}}{\longrightarrow} \;\;  SU(3) \times SU(2)_L \times SU(2)_R \times U(1)_{B-L} \times P_{LR}.
\end{align}
The next such smallest representation is ${\bf 210}$ ($\equiv \phi_{210}$) with 
\begin{align}
\label{eq:210SB422}
SO(10) \times CP \;\;  \stackrel{ \phi_{210}}{\longrightarrow} \;\;  SU(4) \times SU(2)_L \times SU(2)_R \times P_{LR}.
\end{align}

%%%%%%%
\subsection{Fermion masses}
%%%%%%%

We briefly comment on how realistic SM Yukawa couplings can be obtained, leaving the detailed analysis to future work.

\subsubsection{Down-type quarks and charged leptons}
As we discuss in the previous section, Yukawa couplings arise from the exchange of heavy states $X\bar{X}$. We introduce three ${\bf 10}$ ($\equiv X_{10}$)%
\footnote{We may introduce ${\bf 120}$ instead of ${\bf 10}$, but the perturbativity is lost around the unification scale unless the scale $v'$ is high.}
with the following interactions,
\begin{align}
\label{eq:de}
{\cal L }_{d,e} =
 \, (\psi_{16} \, x_{d,e} \, X_{10}) \,\phi_{H} \, {\cal O}_G +  M_{X_{d,e}} \, X_{10}^2 \, {\cal O}_G + {\rm h.c.},
\end{align}
where ${\cal O}_G$ denotes possible insertions of $SO(10)$ symmetry breaking fields.%

After integrating out $X_{10}$, the dimension 5 operators of Eq.~(\ref{eq:high_dim}) for down-type quarks and charged leptons are obtained.
Without ${\cal O}_G$, down-type quarks and charged leptons of the same generation have identical Higgs couplings at $M_G$, in contradiction with the observed spectrum. The observed masses can result, for example, by non-degeneracies in $X_{10}$ from $SO(10)$ symmetry breaking.

\subsubsection{Up-type quarks and neutrinos}
For the up Yukawa couplings we introduce ${\bf 54}$ and/or ${\bf 45}$ ($\equiv X_{54},X_{45}$). We first discuss the case with ${\bf 54}$. The following interaction leads to the up Yukawa couplings,
\begin{align}
{\cal L }_{u_{\bf 54}} =\, (\psi_{16} \, x_u \, X_{54}) \, \phi_{H}^* \, {\cal O}_G + M_{X_u} \, X_{54}^2 \,{\cal O}_G.
\end{align}
When the Higgs is embedded into ${\bf 16}$, the first term vanishes without ${\cal O}_G$.
The $O(1)$ top Yukawa coupling requires that $\vev{{\cal O}_G}=O(1)$, and hence this operator must be generated by an exchange of particles with masses near $M_G$.
On the other hand, when the Higgs is embedded into ${\bf 144}$, the first term does not require the insertion of ${\cal O}_G$.

$X_{54}$ contains an $SU(3)_c\times SU(2)_L \times SU(2)_R \times U(1)_{B-L}$ singlet. Integrating out this singlet generates
\begin{align}
{\cal L}_{\nu_{\bf 54}} = \, \frac{x_u^2}{M_{X_u}} (\ell H^* + \ell' H^{'*})^2.
\end{align}
After $H$ and $H'$ condense, this interaction gives a mass only to a linear combination of $\nu$ and $\nu'$, which is dominantly $\nu'$.
The SM left-handed neutrino masses are not related to the SM up-type quark masses.

Next we consider the case with ${\bf 45}$. The up Yukawa couplings are obtained from 
\begin{align}
{\cal L }_{u_{\bf 45}} =\, (\psi_{16} \, x_u \, X_{45}) \, \phi_{H}^* \,{\cal O}_G + M_{X_u} \, X_{45}^2 \,{\cal O}_G.
\end{align}
In contrast to the case with ${\bf 54}$, the first term is non-zero even if the Higgs is embedded into ${\bf 16}$ and without insertion of the $SO(10)$ symmetry breaking fields.

The interaction, however, in general generates left-handed neutrino masses via the exchange of the $SU(2)_L$ triplet in $X_{45}$. Without any $SO(10)$ symmetry breaking from $\vev{{\cal O}_G}$, the left-handed neutrino masses at tree-level are $m_{u,c,t} \times v / v'$, which are too large and hierarchical.
This can be avoided by 1) generating the up Yukawa couplings predominantly from ${\bf 54}$ or 2) picking up $SO(10)$ symmetry breaking effects from $\vev{{\cal O}_G}$.  For the latter option, for example, consider the dimension-5 operator
\begin{align}
\label{eq:missing yukawa}
{\cal L}_{u_{\bf 45}}^{(d=5)} = \frac{1}{M} (\psi_{16} \, x_u \, \Gamma^{mn}  X_{45}^{m k}) \, \phi_{H}^* \phi_{45}^{kn},
\end{align}
where $m,n,k=1 \cdots 10$ are the vector indices of $SO(10)$. Because of the missing vev structure of $\phi_{45}$, this operator does not give a Yukawa coupling to the $SU(2)_L$ triplet, and hence leaves the SM neutrinos massless while giving up-type quark masses.
The particular $SO(10)$ structure in Eq.~(\ref{eq:missing yukawa}) can be obtained by introducing a pair of ${\bf 45}$s, $Y_{45}$ and $\bar{Y}_{45}$, and the interaction
\begin{align}
{\cal L} \sim \, (\psi_{16} Y_{45}) \phi_{H}^* + (X_{45}  \bar{Y}_{45})  \phi_{45}+ M Y_{45}\bar{Y}_{45}
\end{align}
with $(\psi_{16} X_{45}) \phi_{H}^*$ suppressed by some parity.

\subsubsection{CKM phase}
The strong CP problem is solved by imposing $SO(10) \times CP$ symmetry.  An $SO(10)$ symmetry breaking field, odd under both $CP$ and the $C_{\rm LR}$ parity contained in $SO(10)$, leaves the product, $P_{\rm LR}$, unbroken.  To obtain the CKM phase, the up or down Yukawa couplings must feel this symmetry breaking, since otherwise they are CP symmetric. One simple possibility is to obtain $M_{X_{d,e}}$ in Eq.~(\ref{eq:de}) via
\begin{align}
{\cal L} = (M^{ij} + i \lambda^{ij} \phi_{45}) \, X_{10,i} X_{10,j}  + {\rm h.c.},
\end{align}
where $M^{ij}$ and $\lambda^{ij}$ are symmetric and anti-symmetry, respectively.
CP requires that $M^{ij}$ and $\lambda^{ij}$ are real. After $\phi_{45}$ obtains its vev, the mass matrix of $X$ is Hermitian, and can explain the CKM phase without introducing the strong CP phase.

%%%%%%%%%%%%%%%%%%
\section{Discussion}
%%%%%%%%%%%%%%%%%%
\label{sec:discussion}

%%%%%%%%%%%%%%%%%%
\subsection{A top-down view}
%%%%%%%%%%%%%%%%%%
\label{subsec:topdown}

We began this paper by introducing a $Z_2$ symmetry and showing that the SM Higgs quartic vanishes at the scale $v'$ where $Z_2$ is spontaneously broken, and we proceeded to explore the consequences.  We argued that there must be a new $SU(2)'$ gauge group and that the Higgs sector for $SU(2) \times SU(2)'$ must be essentially minimal: $H(2,1)+H'(1,2)$.  We constructed the set of minimal theories at $v'$: those based on gauge group 3221 with a prime sector and those based on 3221 or 422 with a left-right symmetry. We demonstrated that if the $Z_2$ included spacetime parity these theories could all solve the strong CP problem, and then we proceeded to explore the possibility of unification.  Here we take an opposite view and start from a particular grand unified theory and show how the SM appears. 

Consider a grand unified theory based on symmetry group $SO(10) \times CP$ and broken by the vev of ${\bf 45}_-$ to $3221 \times P_{LR}$.  Light matter is contained in three ${\bf 16}$, $\psi_i$, and the Higgs surviving below the unified scale, $(H,H')$, lies in a ${\bf 16}$, $\phi_H$, rather than the usual ${\bf 10}$.  Since the Yukawa interaction $\psi_i \psi _j \phi_H$ is not gauge invariant, quark and lepton masses must arise from interactions with other fields $X$.   On integrating out $X$, flavor appears in higher-dimensional operators which contain $SO(10)$ breaking, avoiding certain unacceptable $u:d:e$ mass ratios.  For example, $d/e$ masses can arise from $X_{\bf 10}$, and $u/\nu$ masses from  $X_{\bf 45}$ or  $X_{\bf 54}$.  In the latter case, care must be taken to ensure a decoupling of $u$ and $\nu$ masses.

The intermediate symmetry $3221 \times P_{LR}$ is broken by $\vev{H'}=v'$ to the SM.  Taking $v' \sim 10^{10}$ GeV, leads to {\it both} precision gauge coupling unification and $\lambda_{SM}(v') \sim  0$; there is a successful prediction that correlates the observed gauge couplings with the Higgs and top quark masses (five low-energy observables are given in terms of four parameters: $g_G, M_G, y_t, v'$).  The $H'$ vev converts the higher dimensional flavor operators to SM Yukawa couplings that are Hermitian, including 1-loop radiative corrections, solving the strong CP problem. 
%%%%%%%%%%%%%%%%%%
\subsection{Final comments}
%%%%%%%%%%%%%%%%%%
\label{subsec:topdown}

We have not explored cosmological consequences of the ideas and theories discussed in this paper.  The maximum temperature that the SM sector is reheated to after inflation must be less than $v'$ to avoid domain walls from spontaneous $Z_2$ breaking.  This is still consistent with leptogenesis from $\nu_R$ decay~\cite{Fukugita:1986hr}, although the constraints become more severe from lower $v'$, and $\nu_R$ are present in all the LR models.  Finally, unlike the axion solution of the strong CP problem~\cite{Peccei:1977hh,Peccei:1977ur,Weinberg:1977ma,Wilczek:1977pj}, the simplest theories discussed here do not include an exotic dark matter candidate.  

There are several key measurements for our scenario.  A more precise determination of the top quark mass would better constrain $\mu_c$ and therefore the $Z_2$-breaking scale $v'$.  In LR theories, a significant fraction of the parameter space with precision gauge coupling unification has proton decay at a rate observable in planned experiments~\cite{Abe:2011ts}, especially if the intermediate symmetry is 422.   Finally, LR models have 2-loop contributions to $\bar{\theta}$ that are likely within 1 or 2 orders of magnitude of the current experimental bound from the neutron electric dipole moment and may be observable by future experiments~\cite{Lamoreaux:2009zz,Baker:2016bzc,Tsentalovich:2014mfa}.
Discovery of such a dipole moment would point to these theories over axion theories, and would provide an important new constraint in constructing theories of flavor.

\let\oldaddcontentsline\addcontentsline% Store \addcontentsline
\renewcommand{\addcontentsline}[3]{}% Make \addcontentsline a no-op

\section*{Acknowledgement}
This work was supported in part by the Director, Office of Science, Office of High Energy and Nuclear Physics, of the US Department of Energy under Contract DE-AC02-05CH11231 and by the National Science Foundation under grants PHY-1316783 and PHY-1521446. 

\let\addcontentsline\oldaddcontentsline% Restore \addcontentsline

\appendix
%%%%%%%%%%%%%%%%
\section{Constraints on the $SU(2)\times SU(2)'$ symmetry breaking sector}
%%%%%%%%%%%%%%%%
\label{sec:Higgs sector}

In this appendix we show that our understanding of $\lambda_{SM}(v')=0$ arising from a $Z_2$ symmetry imposes very severe constraints on the $SU(2)_L\times SU(2)'$ symmetry breaking sector.
Our mechanism relies on their being a single relevant parameter in the potential that breaks an accidental $SU(4)$ symmetry.  
From (\ref{eq:potential}), $V/\lambda$ depends on two parameters: $m^2/\lambda = v'^2$ which has dimensions and
$\lambda'/\lambda=x$ which is dimensionless. Hence, the low energy potential for $H$ must take the form 
\begin{align}
\label{eq:potentialLE2}
\frac{V_{LE}(H)}{\lambda} = f(x) \; v'^2  \; H^\dagger H + g(x) (H^\dagger H)^2 .
\end{align}
At the $SU(4)$ invariant point $x=0$ both $f$ and $g$ must vanish.  Hence the fine-tune on the quadratic term for  $v \ll v'$ {\it necessarily} forces $\lambda_{SM} \ll 1$.  (From explicit computation, we find that at tree level $f(x) = x$ and $g(x) = x(1+x/2)$, corresponding to (\ref{eq:potentialLE})).   

Now consider additional scalar fields, $\chi$, that acquire vevs at scale $v'$, so that the $Z_2$ invariant potential of the full theory $V(H,H', \chi)$ has additional dimensionless couplings that break the $SU(4)$ invariance.   The low energy effective theory for $H$ now arises from more than one $SU(4)$ breaking parameters, $x_i$, and takes the form of (\ref{eq:potentialLE2}), with $f(x) \rightarrow f(x_i)$ and $g(x) \rightarrow g(x_i)$.  The fine-tune $f(x_i) \ll 1$, necessary for $v \ll v'$, restricts ${x_i}$ to lie on a surface in the parameter space.  Although the surface passes through the origin, where $SU(4)$ is restored, a generic point on the surface has large $SU(4)$ breaking and hence an order unity value for $g(x_i)$. To understand the near vanishing of $\lambda_{SM}$ at $v'$, the scalar potential of the full theory must not have any $SU(4)$ breaking couplings beyond $\lambda'$ of (\ref{eq:potential}).

There are two possible such additions to the scalar sector that contribute to $SU(2)'$ symmetry breaking at scale $v'$ 
\begin{itemize}
\item Multiple copies of doublets $H_a(2,1) \leftrightarrow H'_a(1,2)$, leading to $SU(4)$ violating via
\begin{align}
\label{eq:DeltaV1}
\Delta V(H_a, H'_a) = \lambda_{ab} \, H_a^\dagger H_b \; H'^\dagger_a H'_b + h.c.
\end{align}
giving multiple contributions to $f(x_i)$ and $g(x_i)$.
\item $\chi(R,1) \leftrightarrow \chi'(1,R)$, with $R \neq 2$, leading to the $SU(4)$ violating interaction
\begin{align}
\label{eq:DeltaV2}
\Delta V(H, H', \chi) = \lambda_{\chi} \,(H^\dagger H - H'^\dagger H')(\chi^\dagger \chi - \chi'^\dagger \chi')
\end{align}
giving additional contributions to $f(x_i)$ and $g(x_i)$ when $\chi'$ acquires a vev. 
\end{itemize}
Additional scalars at scale $v'$ are possible if they don't acquire a vev,  We do not consider them as they introduce extra fine-tuning to the theory.

Our mechanism for understanding why $\lambda_{SM}(v')$ is very small requires that the low energy field that breaks $SU(2)_L$ arises predominantly from $H(2,1)$.  For example, consider a bi-doublet with $\phi(2,\bar{2}) \leftrightarrow \phi^\dagger(\bar{2},2)$.  This leads to the $SU(4)$ breaking interaction
\begin{align}
\label{eq:DeltaV3}
\Delta V(H, H', \phi) = \lambda_{\phi} v'\,(H^\dagger \phi H' + H'^\dagger \phi^\dagger H)
\end{align}
with $\lambda_{\phi}$ real. The $H'$ vev now leads to mass mixing between $H$ and $\phi$, so that an additional $SU(4)$ breaking parameters appear in $H/\phi$ mass terms.  Fine-tuning one combination to be light no longer guarantees that $g(x)$ vanishes.
Requiring $\lambda_{SM}(v') \lsim 10^{-2}$ implies that the light SM Higgs contains no more than 10\% of $\phi$. It is possible that additional scalars are in the theory at scale $v$, but they cannot substantially take part in electroweak symmetry breaking, and we do not add them as they imply further fine-tuning.
 
An explanation of the observed value of the Higgs boson mass, leads us to consider an enlarged gauge symmetry, $SU(2) \rightarrow SU(2)_L \times SU(2)'$, broken by $H(2,1) + H'(1,2)$.

%%%%%%%%%%%%%%%%%%%%%%
\section{$SU(4)\times SU(2)_L\times SU(2)_R$ breaking at scale $v'$}
%%%%%%%%%%%%%%%%%%%%%%
\label{sec:422 at vR}

In this appendix we discuss the breaking of $SU(4)\times SU(2)_L\times SU(2)_R$ to the SM gauge group by the vev of $H'$. We denote the $({\bf 4}, {\bf 2}, {\bf 1})$ and $({\bf \bar{4}}, {\bf 1}, {\bf 2})$  Higgses as $\Phi^a_\alpha$ and $\Phi'_{a,\alpha'}$, respectively. Here $a$, $\alpha$, $\alpha'$ are the $SU(4)$, $SU(2)_L$, $SU(2)_R$ indices, respectively.
We decompose $\Phi$ and $\Phi'$ as
\begin{align}
\Phi =
\begin{pmatrix}
d & u \\
0 & h
\end{pmatrix}, \hspace{0.5in}
\Phi' =
\begin{pmatrix}
\bar{d} & \bar{u} \\
0 & h'
\end{pmatrix}
\end{align}
where we have eliminated the $(4,1)$ components by $SU(2)$ rotations.
The vacuum we consider is $u=d=\bar{u}=\bar{d}=h=0$ and $h'\neq 0$, with parameters chosen so that the mass of $h$ nearly vanishes.

The $SU(4)\times SU(2)_L\times SU(2)_R\times P_{LR}$ invariant potential is given in general by
\begin{align}
V = & - m^2 \left( |\Phi|^2 + |\Phi'|^2 \right) + \frac{\lambda}{2} \left( |\Phi|^4 + |\Phi'|^4    \right) + y |\Phi|^2 |\Phi'|^2 + \frac{k}{2} \left(  \Phi^a_\alpha \Phi^{b \alpha}  \Phi_{a\beta}^{*} \Phi_{b}^{*\beta} + \Phi^{'a}_{\alpha'} \Phi^{'b \alpha'}  \Phi_{a\beta'}^{'*} \Phi_{b}^{'*\beta'}  \right) \nonumber \\
 & + \left( \frac{l}{2} \Phi^a_\alpha \Phi^{b \alpha} \Phi_a^{'\beta'} \Phi'_{b \beta'} + {\rm h.c.} \right) + g \Phi^a_\alpha \Phi^{*\alpha}_b \Phi'_{a\beta'} \Phi^{'*\beta'},
\end{align}
where $|\Phi|^2 \equiv \Phi^{a\alpha} \Phi^*_{a \alpha}$. 
The quadratic terms and quartic terms relevant for vacuum stability and for quantum corrections to the SM Higgs quartic coupling are
\begin{align}
V =& - m^2 \left( |h|^2 + |h'^2| + |u|^2 + |\bar{u}|^2 + |d|^2 + |\bar{d}|^2  \right) + \frac{\lambda}{2} \left( \left(|h|^2  + |u|^2 + |d|^2 \right)^2 + \left(|h'|^2  + |\bar{u}|^2 + |\bar{d}|^2 \right)^2 \right) \nonumber \\
& + y (|h|^2  + |u|^2 + |d|^2)(|h'|^2  + |\bar{u}|^2 + |\bar{d}|^2) - g \left( |h|^2|h'|^2 + h^*h'^* \bar{u}^t u + hh' \bar{u}^\dag u^*\right) \nonumber \\
 &  + k \left( |h|^2|d|^2+ |h'|^2 |\bar{d}|^2  \right)+ \left( l \, h h' \bar{d}^t d +{\rm h.c.} \right).
\end{align}

Let us first investigate the conditions for the mass of $h$ to vanish and for vacuum stability. The potential of $h$ and $h'$ is
\begin{align}
V(h,h') = - m^2\left( |h|^2 + |h'|^2 \right) + \frac{\lambda}{2}\left( |h|^4 + |h'|^4\right) + (y-g)|h|^2|h'|^2. 
\end{align}
At the tree level, the vev of $h'$ is $\vev{h'}= m / \sqrt{\lambda} \simeq v'$. The condition for the $h$ mass to vanish is $y-g = \lambda$.
For this value of $y$, the mass-squareds of colored particles are given by
\begin{align}
\left( m_u^2,m_{\bar{u}}^2, m_d^2, m_{\bar{d}}^2 \right) =  \vev{h'}^2 \left( g,0,g,k\right).
\end{align}
$\bar{u}$ is the Nambu-Goldstone boson eaten by the colored gauge boson.
The stability of the vacuum requires that $g,k>0$.

Next we discuss quantum correction to the SM Higgs quartic coupling. The couplings of $h$ and $h'$ with the colored Higgses explicitly break the $SU(4)$ symmetry on $(H,H')$ ($SU(2)$ on $(h,h')$ after eliminating the $(4,1)$ components), and hence these couplings give quantum corrections to the SM Higgs quartic coupling at scale $v'$.
Since we are interested in the one-loop correction, we may use the tree-level relation $y= \lambda + g $ in the coupling of $h$ and $h'$ with the colored Higgses. For this value, the coupling between $h,h'$ and $u,\bar{u}$ is of the form
\begin{align}
\lambda \left(|h|^2 + |h'|^2 \right) \left( |u|^2 + |\bar{u}|^2 \right) + g |h \bar{u}^* - h'^* u|^2,
\end{align}
which is $SU(2)$ invariant. Thus, quantum corrections from $u$, $\bar{u}$ loops do not generate a SM Higgs quartic coupling at the one-loop level.
The masses of $d$ and $\bar{d}$, which mix with each other for generic values of $h$ and $h'$, are given by
\begin{align}
m^2_{\pm} = - m^2 + \frac{1}{2} ( 2\lambda + k + g) (|h|^2 + |h'|^2) \pm \frac{1}{2} \sqrt{ (k-g)^2 (|h|^2 + |h'|^2)^2 + 4 |h|^2 |h'|^2 (|l|^2 - (k-g)^2)  }.
\end{align}

The one-loop corrected effective potential of $h$ and $h'$ is given by
\begin{align}
V_{\rm eff} (h,h') = &  m^2\left( |h|^2 + |h'|^2 \right) + \frac{\lambda}{2}\left( |h|^4 + |h'|^4\right) + (y-g)|h|^2|h'|^2 \nonumber  \\
 & +  \frac{6}{64\pi^2} m_+^4 \; {\rm ln}\frac{m_+^2}{m^2} +  \frac{6}{64\pi^2} m_-^4 \; {\rm ln}\frac{m_-^2}{m^2}.
\end{align}
Here we take the renormalization scale to be $m$. We minimize this potential with respect to $h'$, obtain the condition that the mass of $h$ vanishes at the minimum, and find that the SM Higgs quartic coupling at $v'$ is
\begin{align}
\lambda_{\rm SM}(v') = \frac{3}{32 \pi^2} |l|^2 \; f \left( \frac{g}{|l|} , \frac{k}{|l|} \right), \\
f(x,y) =  \frac{(1 - (x-y)^2)^2}{(x-y)^3}  \left( 2 \left(x-y\right) + \left(x +y \right) {\rm ln} \frac{y}{x}  \right).
\end{align}
with the normalization $V = \lambda_{\rm SM} |H|^4$. The function $f(x,y)$ is always negative, and for $O(1)$ $(x,y)$ its typical size is $O(0.1)$. Thus, for $O(1)$ couplings $l,g,k$, the $\lambda_{\rm SM}(v')$ is negative and typically $O(10^{-3})$.  However, there are $O(1)$ couplings that give $f$ a size $O(1)$, allowing $v'$ as large as $10^{14}$ GeV, even for central values of the top quark mass.

%%%%%%%%%%%%%%%%%%%%
\section{Quantum corrections to the strong CP phase}
%%%%%%%%%%%%%%%%%%%%
\label{sec:correction}

In this appendix, we estimate the quantum corrections to the strong CP phase in Model A.
The Yukawa couplings and the mass term of heavy states $X\bar{X}$ are given by
\begin{align}
\label{eq:interaction}
{\cal L} = & (q^i \, x^u_{ia} \bar{X}^a_u) H^\dag + (q'^i \, x^{'u}_{ia} X^a_u) H^{'\dag} + M^u_{a} X^a_u\bar{X}^a_u \nonumber \\
&  +  (q^i \, x^d_{ia} \bar{X}^a_d) H + (q'^i \, x^{'d}_{ia} X^a_d) H^{'} + M^d_{a} X^a_d\bar{X}^a_d +  {\rm h.c.}
\end{align}
Here  we choose $X$ to transform as (1,1) under $(2_L,2_R)$. The alternative of $X$(2,2) gives Yukawa interactions $q \bar{X}H' + q' XH$, but the estimation is essentially the same.
Ref.~\cite{Babu:1989rb} shows that there is no one-loop quantum correction to the strong CP phase. As we will see two-loop corrections do not vanish and may give an observable neutron electric dipole moment in future experiments.
We do not consider theories having $X$ states with both charges.

We can assign the spurious symmetry shown in Table~\ref{tab:spurious symmetry}.
The parity requires that $x'_{ij}=x_{ij}^*$ and $M_a$ are real. To make use of the spurious symmetry, we keep $x'$ as an independent coupling and $M_a$ as a complex number until we evaluate the reality of the quantum correction.

\begin{table}[tb]
\caption{Spurious symmetry of the interaction in Eq.~(\ref{eq:interaction}).}
\begin{center}
\begin{tabular}{|c||c|c|c|c|c|c|c|c|c|c|c|c|c|c|} \hline
				&$q^i$	&$q'^j$	&$\bar{X}_u^a$	&$X_u^b$	&$\bar{X}_d^c$		&$X_d^e$	&$H$	&$H'$	&$x^u_{ia}$	&$x^{'u}_{jb}$	&$x^d_{ic}$	&$x^{'d}_{je}$	&$M^u_a$	&$M^d_c$ \\ \hline
$SU(3)_q$		&${\bf 3}$	&			&			&		&				&		&		&		&${\bf \bar{3}}$	&			&${\bf \bar{3}}$	&			&			&	\\
$SU(3)_{\bar{q}'}$	&		&${\bf 3}$		&			&		&				&		&		&		&			&${\bf \bar{3}}$	&			&${\bf \bar{3}}$	&			&	\\
$[U(1)_{\bar{u}}]^3$	&		&			&$1$			&		&				&		&		&		&$-1$		&			&			&			&$-1$		&	\\
$[U(1)_{u}]^3$		&		&			&			&$1$		&				&		&		&		&			&$-1$		&			&			&$-1$		&	\\
$[U(1)_{\bar{d}}]^3$	&		&			&			&		&$1$				&		&		&		&			&			&$-1$		&			&			&$-1$ \\
$[U(1)_{d}]^3$		&		&			&			&		&				&$1$		&		&		&			&			&			&$-1$	 	&			&$-1$	\\
$U(1)_H$			&		&			&			&		&				&		& $+1$	&		& $+1$		&			& $-1$		&			&			&\\
$U(1)_{H'}$		&		&			&			&		&				&		&		& $+1$	&			& $+1$		&			& $-1$		&			&\\ \hline
\end{tabular}
\end{center}
\label{tab:spurious symmetry}
\end{table}%

After integrating out $X \bar{X}$, the following higher-dimensional operators are generated,
\begin{align}
\label{eq:non hermitian mass}
{\cal L}_{\rm eff} = q^i \left(  c^u_{ij} f^u(|H'|^2) + c^{u\dagger}_{ij} f^u(|H|^2)  \right) q^{'j} H^\dag H^{\dag'} + q^i \left(  c^d_{ij} f^d(|H'|^2) + c^{d\dagger}_{ij} f^d(|H|^2)  \right) q^{'j} H H' + {\rm h.c.}.
\end{align}
so that the quark Yukawa matrices are no longer Hermitian. The correction to the strong CP phase is given by
\begin{align}
\label{eq:thetagen}
\Delta \theta = {\rm Im} \left[(y^{u})^{-1}_{ji} c^u_{ij} v' f^u (v^{'2})\right] + {\rm Im} \left[(y^{d})^{-1}_{ji} c^d_{ij} v' f^d (v^{'2})\right],
\end{align}
where the Yukawa matrix and its inverse are given by
\begin{align}
\label{eq:yyinv}
y_{ij} = x_{ia} \frac{v'}{M_a} x'^T_{aj} = x_{ia} \frac{v'}{M_a} x^\dagger_{aj} ,~~y^{-1}_{ij} = x'^{T-1}_{ia} \frac{M_a}{v'} x^{-1}_{aj} = x^{\dagger -1}_{ia} \frac{M_a}{v'} x^{-1}_{aj}.
\end{align}

The spurious symmetry fixes the coefficient $c_{ij}^u$ at $O({x^2})$ to be of the form
\begin{align}
c_{ij}^{u(2,0)} \sim x^u_{ia} x^{'u}_{ja}M_a^* F_a = x^u_{ia} x^{u\dagger}_{aj} M_a F_a ,
\end{align}
where $F_a$ is a real function of $|M_a|$, and in the final expression we have imposed $Z_2$. One can easily verify that $c_{ij}^{u(2,0)}$ is Hermitian and does not contribute to $\Delta \theta$. Similarly, the coefficient $c_{ij}^d$ at $O({x^2})$ does not generate $\Delta \theta$.

The coefficient $c_{ij}^u$ at $O({(x^u)^4})$ is of the form
\begin{align}
c_{ij}^{u(4,0)} &\sim x^u_{ia} x^{'u}_{jb} x^u_{mb}  x^{u*}_{ma} M_b^*F_{ab} + x^u_{ia} x^{'u}_{jb} x^{'u}_{ma}  x^{'u*}_{mb} M_a^* F'_{ab} + x^u_{ia} x^{'u}_{ja} M_a^* O(|x^u|^2) \nonumber \\
 &=   x^u_{ia} x^{u\dagger}_{bj} x^u_{mb}  x^{u\dagger}_{am} \tilde{F}_{ab}  + x^u_{ia} x^{u\dagger}_{aj} M_a O(|x^u|^2),
\end{align}
where $\tilde{F}_{ab} = M_b^*F_{ab}  +  M_a^*F'_{ab}$.
The second term in the last line is of the same form as $c_{ij}^{u(2,0)}$ and does not contribute to $\Delta \theta$. The first term, contracted with $(y^u)^{-1}$, is
\begin{align}
(y^{u})^{-1}_{ji} c^{u(4,0)}_{ij} = (x^\dagger)^{-1}_{jn} M_a (x)^{-1}_{ni} x^u_{ia} x^{u\dagger}_{bj} x^u_{mb}  x^{u\dagger}_{am} \tilde{F}_{ab}  = M_a x^u_{ma} x^{u\dagger}_{am} \tilde{F}_{aa},
\end{align}
which is real. We obtain the same conclusion for $c_{ij}^d$ in $O({(x^d)^4})$.

The first non-zero contribution arise from $O((x^u)^2 (x^d)^2 )$,
\begin{align}
c_{ij}^{u(2,2)} \sim x^u_{ia} x^{'d}_{jb} x^{'u}_{ka} x^{'*d}_{kb} M^{u*}_a G_{ab} = x^u_{ia} x^{d\dagger}_{bj} x^{u\dagger}_{ak} x^{d}_{kb} M^{u}_a G_{ab}.
\end{align}
$c_{ij}$ receives a non-Hermitian contribution if $k \neq i$ or $k \neq j$, which manifests the necessity of the correction involving at least two generations.

%%%%%%%%%%%%%%%%%%%%%%%%%%%%%%%%%%%%%%%%%%%%%%%%%%%%%%%%%%%%%
\begin{figure}[htbp]
\centering
\includegraphics[width=0.7\linewidth]{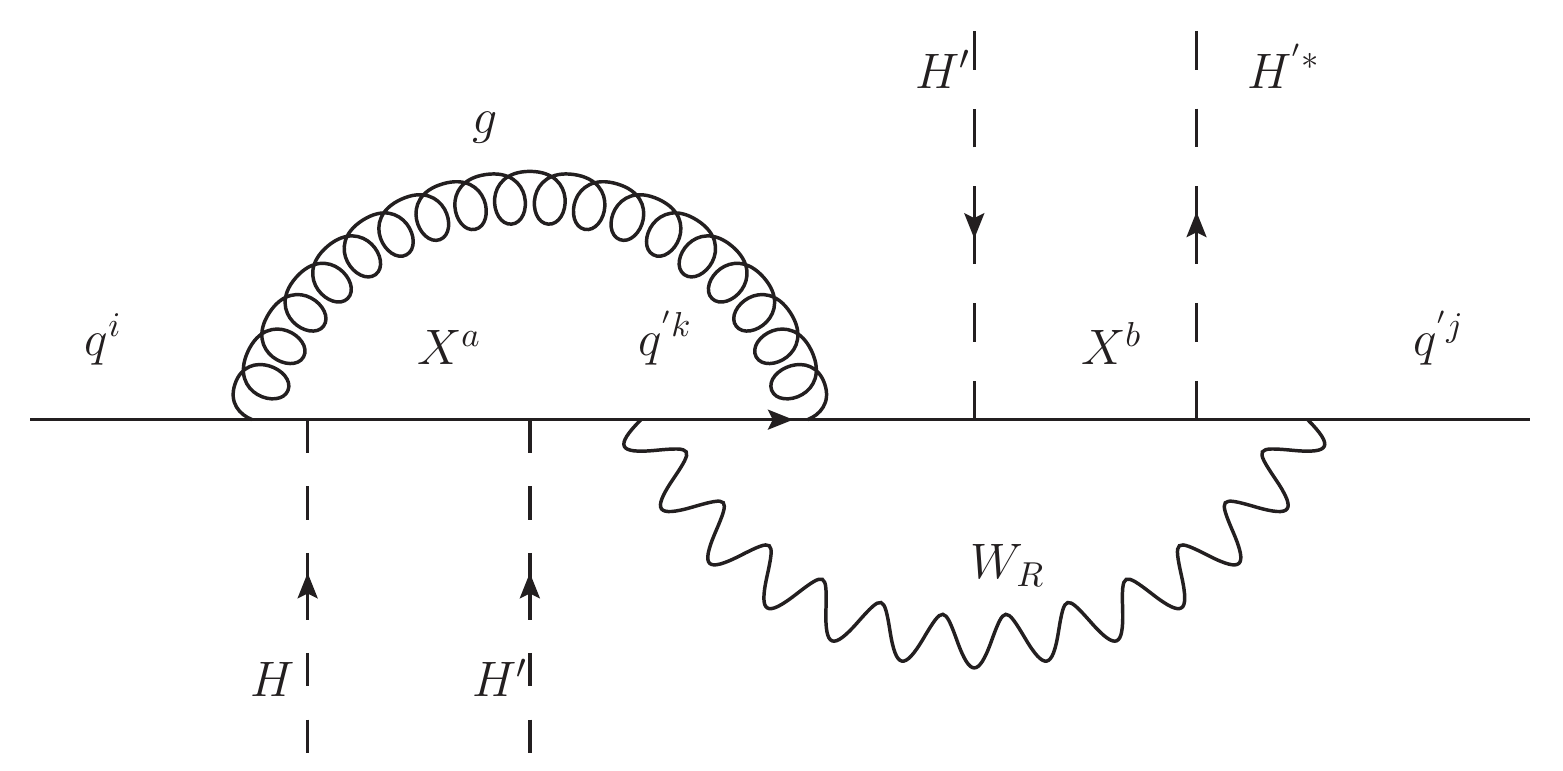}
\caption{\textsl{\small
One of the two-loop corrections to the quark mass which lead to a non-zero strong CP phase.
}}
\label{fig:diagram}
\end{figure}
%%%%%%%%%%%%%%%%%%%%%%%%%%%%%%%%%%%%%%%%%%%%%%%%%%%%%%%%%%%%%%%%%%%%%%%%

One of the diagrams that generates $c_{ij}^{u(2,2)}$ is shown in Fig.~\ref{fig:diagram}. The $W_R$ boson is required to have a $O((x^u)^2 (x^d)^2 )$ insertion, and without the gluon the diagram only renormalizes the wave function of $q'$. The diagram gives
\begin{align}
c_{ij}^{u(2,2)} f^u(\vev{H'}^2) &\simeq \frac{g_2^2 g_3^2}{(16\pi^2)^2} v^{'2} x^u_{ia} x^{d\dagger}_{bj} x^{u\dagger}_{ak} x^{d}_{kb} M^{u}_a G_{ab} \nonumber \\
&\simeq \frac{g_2^2 g_3^2}{(16\pi^2)^2} v^{'2} x^u_{ia} x^{d\dagger}_{bj} x^{u\dagger}_{ak} x^{d}_{kb}  \frac{f_{ab}}{M^u_a (M_b^d)^2},
\end{align}
where we use $G_{ab} \sim (M^u_a M_b^d)^{-2}$ and $f_{ab}$ are $O(1)$ numbers.

The contribution to $\Delta \theta$ from  $c_{ij}^{u(2,2)}$ depends on the structure of $x^u$ and $x^d$, but we can derive a rough upper bound on the correction. 
Using the relation $y_{ij} = v'\sum_kx_{ia} x_{ja}^*/M_a$, we obtain $|x_{ia}x^\dagger_{aj}| \lsim |y_{ij}|M_a /v' $, giving
\begin{align}
v' \, |c_{ij}^{u(2,2)} f^u(|H'|^2)| \; \lesssim \; \frac{g_2^2 g_3^2}{(16\pi^2)^2} \,  |y_{ik}^u||y_{kj}^d| \frac{v'}{M_b^d} \; \lesssim \; \frac{g_2^2 g_3^2}{(16\pi^2)^2} \, |y_{ik}^u||y_{kj}^d|,
\end{align}
where we take the minimal possible value for $M_b^d$, of $v'$. For larger values the correction is chirality-suppressed by $v'/M$. The bound is saturated when the relevant Yukawa couplings are generated by heavy states $X$ with mass near $v'$. 
Using (\ref{eq:thetagen}), the corresponding upper bound on $\Delta \theta$ is given by
\begin{align}
\label{eq:Dtheta}
\Delta \theta_u \, \lesssim & \, \frac{g_2^2 g_3^2}{(16\pi^2)^2} \, (y^u)^{-1}_{ji} \, |y_{ik}^u||y_{kj}^d|, \nonumber \\
\Delta \theta_d \, \lesssim & \, \frac{g_2^2 g_3^2}{(16\pi^2)^2} \, (y^d)^{-1}_{ji} \, |y_{ik}^u||y_{kj}^d|
\end{align}
for the up and down quark masses, respectively.
Assuming that the contributions of  $y^u$ and $y^d$ to the CKM mixing is of the same order, $y^u$, $y^d$, and their inverses are roughly given by
\begin{align}
\label{eq:ynumbers}
y^u \sim
\begin{pmatrix}
y_u			& y_c \theta_{12}	& y_t \theta_{13} \\
y_c \theta_{12}	& y_c			& y_t \theta_{23} \\
y_t \theta_{13}	& y_t \theta_{23}	& y_t
\end{pmatrix} \sim
\begin{pmatrix}
4\times10^{-6}	& 4\times10^{-4}	& 2\times10^{-3} \\
4\times10^{-4}	& 2\times10^{-3}	& 2\times 10^{-2} \\
2\times10^{-3}	& 2\times 10^{-2}	& 0.5
\end{pmatrix}, \nonumber \\
y^d \sim
\begin{pmatrix}
y_d			& y_s \theta_{12}	& y_b \theta_{13} \\
y_s \theta_{12}	& y_s			& y_b \theta_{23} \\
y_b \theta_{13}	& y_b \theta_{23}	& y_b
\end{pmatrix} \sim
\begin{pmatrix}
8\times10^{-6}	& 3\times10^{-5}	& 3\times10^{-5} \\
3\times10^{-5}	& 2\times10^{-4}	& 3\times 10^{-4} \\
3\times10^{-5}	& 3\times 10^{-4}	& 8 \times 10^{-3}
\end{pmatrix}, \nonumber \\
\nonumber \\
(y^u)^{-1} \sim
\begin{pmatrix}
10^4& 10^{3}	& 10^{2} \\
10^3	& 10^2	& 10	 \\
10^2	& 10	& 1
\end{pmatrix},~
(y^d)^{-1} \sim
\begin{pmatrix}
10^5& 10^{5}	& 10^{3} \\
10^5	& 10^4	& 10^3	\\
10^3	& 10^3	& 10^2
\end{pmatrix},
\end{align}

In Table~\ref{tab:correction}, we list the contribution to $\Delta \theta$ for each combination $(i,j,k)$ in Eq.~(\ref{eq:Dtheta}) which exceeds $O(10^{-9})$;
the maximal correction to the strong CP phase is $10^{-6}$.

\begin{table}[htp] 
\caption{Maximal possible corrections to the strong CP phase from $(i,j,k)$ terms in Eq.~(\ref{eq:Dtheta}). With hierarchical $M_a$, the correction is much smaller.}
\begin{center}
\begin{tabular}{|c|c|c|} \hline
 		& $\Delta \theta_u$		& $\Delta \theta_d$ \\ \hline
$10^{-6}$	&					& (2,2,3)			\\
$10^{-7}$	&					& (1,1,3)			\\
$10^{-8}$	&					& (1,1,2), (1,2,3), (1,2,2), (2,3,3) \\
$10^{-9}$	& (1,1,3), (1,2,3), (2,2,3)	& (1,3,3), (2,2,1), (2,3,2),  (3,3,2) \\ \hline
\end{tabular}
\end{center}
\label{tab:correction}
\end{table}%

The maximal value is achieved by assuming that the relevant tree-level Yukawa couplings, $y$ of (\ref{eq:yyinv}), are generated by the exchange of heavy states with masses around $v'$. 
A more natural assumption is that the quark mass hierarchies reflect hierarchies in the masses of the heavy $X$ states. Hence we consider
$v'\sim M_3^u \ll M_3^d \sim M_2^u \ll M_2^d \ll M_1^u\sim M_1^d$, and investigate $c_{ij}^{d(2,2)}$
\begin{align}
\label{eq:cd}
c_{ij}^{d(2,2)} f^d(\vev{H'}^2) \, &\simeq \, \frac{g_2^2 g_3^2}{(16\pi^2)^2}\, x^d_{ia} x^{u\dagger}_{bj} x^{d\dagger}_{ak} x^{u}_{kb} \,  v'^2  \frac{f_{ab}}{M^d_a (M_b^u)^2}.
\end{align}
The factor $1/M^d_a (M_b^u)^2$ is maximized for $b=3$ because of the smallness of $M_3^u$. Then $f_{a3}$ is of the form
\begin{align}
\label{eq:Gk3}
f_{a3} \sim c_0 +  \frac{c_a (M_3^u)^2}{(M_a^d)^2},
\end{align}
where the coefficient $c_0$ does not depend on $M_a^d$. This is because the first term comes from loop momenta around $M_3^u \ll M_a^d$. Taking the leading term, $c_{ij}^{d(2,2)}$ is given by
\begin{align}
c_{ij}^{d(2,2)} f^d(\vev{H'}^2) \, &\simeq \, \frac{g_2^2 g_3^2}{(16\pi^2)^2} \,  y^d_{ik} x^{u\dagger}_{3j} x^{u}_{k3} \, \frac{v'}{(M_3^u)^2}.
\end{align}
This leading contribution, after contraction with $(y^d)^{-1}$, becomes real and does not generate the strong CP phase.
The non-zero contribution comes from the second term in Eq.~(\ref{eq:Gk3}) or the $b\neq 3$ terms in Eq.~(\ref{eq:cd}), which are suppressed at least by $(M_3^u / M_3^d)^2$.

The leading relevant correction to the down Yukawa is thus
\begin{align}
\label{eq:deltayd}
v' \, |c_{ij}^{d(2,2)} f^d(v'^2)|  \; \lesssim \; \frac{g_2^2g_3^2}{(16\pi^2)^2} \, x^d_{i3} x^{d\dagger}_{3k}  x^{u}_{k3} x^{u\dagger}_{3j}  \,  \frac{v'^3}{(M^d_3)^3}  \; \lesssim \; \frac{g_2^2g_3^2}{(16\pi^2)^2} \, y^d_{ik}y^u_{kj} \frac{v'M^u_3}{(M^d_3)^2} \simeq  \frac{g_2^2g_3^2}{(16\pi^2)^2} \, y^d_{ik}y^u_{kj} \frac{y_b^2}{y_t},
\end{align}
where an order unity phase is omitted. The inequalities are saturated when the relevant yukawa couplings dominantly comes from the exchanges of $X_3$. The dominant contribution to $\Delta \theta$ again comes from $(2,2,3)$, and is given by
\begin{align}
\Delta \theta_d \lesssim \frac{g_2^2g_3^2}{(16\pi^2)^2} \, \theta^d_{23}\theta^u_{23} \frac{y_b^3}{y_s},
\end{align}
where $\theta_{23}$ is the mixing angle between the second and the third generations.  Note that left- and right-handed angles are essentially equal due to the leading Hermiticity of the Yukawa matrices.
The CKM mixing matrix elements of the second and the third generations are $V_{cb} = |\theta^d_{23} - \theta^u_{23}|$. A prediction for $\bar{\theta}$, and hence for the neutron electric dipole moment, is therefore model-dependent because (\ref{eq:deltayd}) depends not only on the CKM mixing, but also on the mixing angles $\theta^u_{23}$ and $\theta^d_{23}$.  Nevertheless, a typical expectation is
\begin{align}
\label{eq:thetabarpred}
\bar{\theta}  \; \simeq  \; \frac{g_2^2g_3^2}{(16\pi^2)^2} \, \frac{y_b^3}{y_s} \, V_{cb}^2 \; \simeq \; {\cal O} (10^{-11}).
\end{align}

Quantum corrections to the mass matrix of $X \bar{X}$, if non-Hermitian, also contribute to the strong CP phase. In the effective theory view point, this is contained in the dimension-6 operator of the form (\ref{eq:HHGG}).  However, the resultant contribution to the strong CP phase is suppressed in a manner similar to the Nelson-Barr mechanism~\cite{Nelson:1983zb,Barr:1984qx,Bento:1991ez}, and hence is sub-dominant in comparison with the one we have estimated.

\let\oldaddcontentsline\addcontentsline% Store \addcontentsline
\renewcommand{\addcontentsline}[3]{}% Make \addcontentsline a no-op

\let\addcontentsline\oldaddcontentsline% Restore \addcontentsline
  
\end{document}